\documentclass[10pt,journal,compsoc]{IEEEtran}
\usepackage{cite}
\ifCLASSINFOpdf
  \usepackage[pdftex]{graphicx}
  \DeclareGraphicsExtensions{.pdf,.jpeg,.png}
\else
\fi
\usepackage{lipsum}
\usepackage{xcolor}
\usepackage{cuted}
\usepackage{amsmath}
\usepackage{amssymb}
\usepackage{bbm}
\usepackage{ragged2e}
\usepackage{hyperref}
\usepackage{graphicx}
\definecolor{NavyBlue}{RGB}{35,35,142}
\definecolor{RawSienna}{RGB}{199,97,20}
\hypersetup{
  colorlinks,%
  citecolor=NavyBlue,%
  filecolor=NavyBlue,%
  linkcolor=RawSienna,%
  urlcolor=NavyBlue
}

\usepackage{amsmath}
\usepackage{algpseudocode}
\usepackage{algorithm}
\algnewcommand{\LineComment}[1]{\State \(\triangleright\) #1}
\usepackage{array}
\newcommand{\ours}{\textsc{LaueOT}}
\newcommand{\spot}{\textsc{Spot}}

\newcommand{\T}{\ensuremath{^{\top}}}
\newcommand{\angstrom}{\textup{\AA}}

\DeclareMathOperator*{\argmax}{argmax}
\newcommand{\shorteq}{\mathrel{\mkern0.2mu\mathpalette\shorteq@\relax\mkern0.2mu}}

\begin{document}
\title{Laue Indexing with Optimal Transport}
\author{Tomasz~Kacprzak,
        Stavros~Samothrakitis,
        Camilla~Buhl~Larsen,
        Jaromír~Kopeček,
        Markus~Strobl,
        Efthymios~Polatidis,
        Guillaume~Obozinski
\IEEEcompsocitemizethanks{
\IEEEcompsocthanksitem TK is with Swiss Data Science Center, Paul Scherrer Institute \protect\\
E-mail: tomasz.kacprzak@psi.ch
\IEEEcompsocthanksitem SS, CBL, MS and EP are with the Paul Scherrer Institute
\IEEEcompsocthanksitem JK is with the Institute of Physics, Czech Academy of Sciences
\IEEEcompsocthanksitem GO is with the Swiss Data Science Center, École Polytechnique Fédérale de Lausanne 
}
\thanks{}}

\markboth{}%
{}

\IEEEtitleabstractindextext{%
\begin{abstract}
\justifying
Laue tomography experiments retrieve the positions and orientations of crystal grains in a polycrystalline samples from diffraction patterns recorded at multiple viewing angles.
The use of a broad wavelength spectrum beam can greatly reduce the experimental time, but poses a difficult challenge for the indexing of diffraction peaks in polycrystalline samples;~the information about the wavelength of these Bragg peaks is absent and the diffraction patterns from multiple grains are superimposed.
To date, no algorithms exist capable of indexing samples with more than about 500 grains efficiently.
To address this need we present a novel method: \emph{Laue indexing with Optimal Transport} (\ours).
We create a probabilistic description of the multi-grain indexing problem and propose a solution based on Sinkhorn Expectation-Maximization method, which allows to efficiently find the maximum of the likelihood thanks to the assignments being calculated using Optimal Transport.
This is a non-convex optimization problem, where the orientations and positions of grains are optimized simultaneously with grain-to-spot assignments, while robustly handling the outliers. 
The selection of initial prototype grains to consider in the optimization problem are also calculated within the Optimal Transport framework.
\ours\ can rapidly and effectively index up to 1000 grains on a single large memory GPU within less than 30 minutes.
We demonstrate the performance of \ours\ on simulations with variable numbers of grains, spot position measurement noise levels, and outlier fractions.
The algorithm recovers the correct number of grains even for high noise levels and up to 70\% outliers in our experiments.
We compare the results of indexing with \ours\ to existing algorithms both on synthetic and real neutron diffraction data from well-characterized samples.
The code and test data are available on \href{link}{\url{https://github.com/LaueOT/laueotx}}.
\end{abstract}
}

\maketitle
\IEEEdisplaynontitleabstractindextext
\IEEEpeerreviewmaketitle
\newpage
\ifCLASSOPTIONcompsoc
\IEEEraisesectionheading{\section{Introduction}\label{sec:introduction}}
\else
\section{Introduction}
\label{sec:introduction}
\fi

\IEEEPARstart{L}{aue} and Bragg diffraction are the foundation of modern experiments in crystallography and solid state physics \cite{Lax1951scattering,WelberryWeber2016onehundred}. 
In neutron and X-ray diffraction, the incoming waves are elastically scattered by the atoms in the crystal lattice of the studied sample.
Due to the regular arrangement of the atoms in the crystal lattice, constructive interference occurs and produces strong reflections, also known as Bragg peaks, which can be detected with diffraction detectors covering the relevant angular range.
Bragg's law describes the relation between the incidence angles, the lattice spacing, and the wavelength of the incoming and reflected waves, which defines the diffraction condition.
This enables modern X-ray and neutron based experimental methods to infer the internal structure of crystalline samples, with numerous applications in material science, chemistry, medicine, biology, electronics, and others. 

The process of inferring the orientation of crystal lattices from the recorded Bragg peak positions is a challenging task: multiple methods have been proposed to solve variants of this problem specific to different types of experimental setups (see \cite{Morawiec2022review} for review). 
One of the most difficult problems in Laue diffraction is indexing of polycrystalline samples, which contain multiple crystal grains with different orientations \cite{Sorensen2012multigrain}, where many overlapping spot patterns are recorded simultaneously. 
The three-dimensional macroscopic structures of samples can be inferred using tomographic techniques, where projections from multiple sample rotations are obtained during the experiment.
The majority of the crystallographic studies use a monochromatic beam or the time-of-flight approach to resolve the wavelength relative to a diffraction peak measured. 
This enables immediate identification of the Miller indices corresponding to the recorded diffraction spot, as the Bragg condition is satisfied only for a single incidence angle between the crystal lattice plane and the incoming beam for a specific wavelength. 
While this simplifies the indexing process, this type of experiment can be time consuming, as monochromatic sources tend have lower overall intensity.
The use of white beams enables faster experiments, but leads to a more complicated inference problem, as an observed spot can stem from almost any crystal lattice plane at any orientation.
This creates a mixed combinatorial and continuous problem, where the peaks-to-grain assignments have to be found jointly with the corresponding grain orientations and positions.
Moreover, the number of grains in the sample is not known a priori and must also be inferred.
Thus, the indexing of white beam polycrystalline patterns is one of the most challenging inference problems among crystallographic experiments.

From an experimental perspective, the last two decades have seen progress in three-dimensional (3D) electron- 
\cite{groeber20063d, zaefferer2008three, liu2011three, konijnenberg2015assessment, stechmann20163, smeets2018serial, zhu2022five} 
and X-ray-based 
\cite{poulsen2001three, larson2002three, jensen2006x, ludwig2008x, johnson2008x, ludwig2009three, ludwig2009new, lienert2011high, jensen2012three, mcdonald2015non} 
grain indexing and reconstruction methods, significantly improving experimental capabilities. 
These methods provide valuable insights into the internal structure of samples and have yielded promising results in grain orientation mapping. 
Despite their impressive spatial resolution, ranging from a few hundred nanometers to micrometers, these techniques are not without limitations. For example, 3D electron backscatter diffraction (3D EBSD) is a destructive method that necessitates serial sectioning of samples. 
Moreover, the penetration of both electrons and X-rays into metallic specimens is constrained, limiting the volume of the sample that can be probed. 
On the other hand, neutron-based methods \cite{Raventos2019laue3dndt, samothrakitis2022microstructural, peetermans2013simultaneous, woracek20143d, peetermans2014cold, sato2017inverse, cereser2017time, woracek2018diffraction, samothrakitis2020grain} 
have emerged as a potential solution owing to their superior penetration ability in various materials. 
These non-destructive techniques enable investigations into large volumes within the bulk of samples. 
Among the recently developed neutron diffraction-based tomography techniques for 3D grain mapping, Laue three-dimensional neutron diffraction tomography (Laue 3DNDT) \cite{Raventos2019laue3dndt} stands out as a particularly promising approach. 
It employs a polychromatic neutron beam, for multi-grain indexing \cite{samothrakitis2022microstructural} and morphology reconstruction \cite{samothrakitis2020grain}, effectively reducing the experimental time while maximizing the information obtained from the sample.  

With multiple solutions proposed for solving this types of problems \cite{Schmidt2014grainspotter,Gevorkov2020pinkindexer,Gevorkov2019xgandalf,Kalinowski2011laueutil,Schlitt2012dbscan,Wejdemann2016dirax,PurushottamRajPurohit2022lauenn,Raventos2019laue3dndt}, only few are directly applicable to wide-beam polycrystalline indexing problem.
Recently, \cite{Raventos2019laue3dndt} proposed a forward-fitting algorithm that achieved good results for samples with up to around 500 grains \cite{samothrakitis2022microstructural}.
However, the method is greedy, incurs in a long runtime, and it is not amenable for parallelization.
With upcoming improvements in experimental techniques and instrumentation in white beam diffraction imaging with neutrons \cite{samothrakitis2023calibr}, there is a clear need for designing novel algorithms for this problem, which allow both for rapid indexing and analyzing samples with orders of magnitude more grains.

The problem of finding orientations and positions of the grains can be approached from the computer vision perspective; the task is in similar to the structure-from-motion \cite{Schonberger2016structure}, camera localization in multi-view geometry (see \cite{HartleyZisserman2004multiview} for overview), as well as registration of point clouds \cite{Huang2021comprehensive}.
The task of spot-to-grain assignment and selection of prototype grains can both be tackled using recent developments in optimal transport (OT), an area of active development \cite{Peyre2019computational}.
In particular, the recently proposed \emph{Sinkhorn Expectation Maximization} method \cite{Mena2020sinkhorn} demonstrated the deep relationship between the OT and the Expectation-Maximization framework, which is commonly used for solving probabilistic models with latent variables.
Regarding methods for finding best-fit parameters, extensive literature exists for convex optimization methods in similar computer vision problems, with coordinate descent enjoying advantage in convergence times in many cases \cite{Wright2015coordinate}.

Utilizing these ideas, we propose a novel technique for indexing of polycrystalline samples from white beam experiments, which we call \ours\ (Laue indexing with Optimal Transport).
We propose a joint inverse problem formulation for the assignment of spots to grains, and the determination of each grain's position and orientation. 
Given an assignment of spots to grains, our formulation benefits from closed-form updates for finding the orientation and grain position, and relies on optimal transport for finding the assignments of spots to grains given candidate grain orientations and positions.
A critical initial step of selecting prototype grains is also handled using the OT framework \cite{Gurumoorthy2021spot}.
This inverse problem formulation enables very fast optimization using GPU-batch solver with convergence guarantees.
The optimal transport methods enable solving for grain parameters and spot assignments \emph{jointly}, which increases the precision of the estimated parameters, the recall of grains found, and the number of correctly assigned spots.

The algorithm has two pre-processing steps: 
(i) creating a set of plausible grain candidates by performing a coarse single-gain fitting in the orientation and position space, and
(ii) selecting the prototype grains using the optimal transport framework, following \cite{Gurumoorthy2021spot}. 
The main solver performs multi-grain fitting, where it jointly optimizes for parameters of the grains and spot-to-grain assignments using the optimal transport framework.
The key aspect of the problem is the treatment of outlier spots in the data and unmatched spots in the model. 
In \ours, we propose a novel probabilistic outlier modeling tailored to the Laue problem.
We compare it to other methods in literature for treatment of outliers: unbalanced and partial OT \cite{Chizat2018scaling,Benamou2015iterative}.

\section{Previous work}
\label{sec:literature}

Multiple Bragg indexing algorithms have been proposed \cite{Morawiec2022review}.
The Hough transform approaches \cite{Sharma2012fast2,Schmidt2014grainspotter} are one of the most commonly used in practice. 
These algorithms use 3D histograms in the orientation space, and each spot ``votes'' for the plausible voxels.
The set of resulting grains corresponds to the voxels with the most votes.
This procedure is then typically followed by a minimizer-based refinement to find the spatial positions of the grains inside the sample.
This method is particularly effective for monochromatic beams, as each spot votes for relatively small number of rotations.
Recently, an extension of this method for pink beams was proposed by \cite{Gevorkov2020pinkindexer}, where \mbox{10-20\%} divergence in the beam spectrum is handled well by the algorithm.
The Hough transform-based methods can be limited by the resolution of voxels in the orientation space, especially that for wider beams the number of votes given by each spot is large, which limits the use of sparse representations of the 3D histogram.
Moreover, most of such methods do not include grain position estimation and use post-processing steps to find it.

To address these problems and enable wide-beam analysis, a forward-fitting method \textsc{Laue3DNDT} was recently proposed \cite{Raventos2019laue3dndt}.
This method first performs an exhaustive search in the space of orientations, keeping the grain positions fixed at the center of the sample.
For each candidate, it predicts the model spots and assigns them to the measurements using a nearest neighbor method.
Then, the loss is calculated as the median distance between spot pairs. 
If the loss is smaller than some threshold, the gradient-based optimization is performed to find the position and orientation.
If certain criteria are met, the candidate is accepted and the assigned spots are removed from further analysis.
The algorithm ends when the number of remaining spots is sufficiently low.
The drawback of this method is its long runtime, which is dominated by slow downhill optimizer.
The greedy nature of the algorithm excludes efficient parallelization.

Recently, machine learning approaches have been proposed to tackle the problem of indexing. 
In \cite{PurushottamRajPurohit2022lauenn}, a fast neural network was created to output grain orientations from X-ray experiments.
This method calculates features for each spot and then passes them through a neural network. 
The feature used is a histogram of angular distances to neighboring spots within a certain radius. 
The method is trained before or during the experiments on the simulations of the corresponding crystal lattice in question.
It reports good performance and practically instant results.
While not tested on polycrystalline data, it could suffer difficulties with interpreting multiple overlapping spot patterns for samples with very large ($>$1000) number of grains.

There exists rich literature in computer vision (CV) and optimal transport (OT) that is relevant to this problem.
The structure-from-motion (SfM) algorithms tackle the problem of reconstructing 3D scene from a series of 2D images, each taken at a different camera position.
The inverse-problem formulation with Expectation-Maximization has been successful \cite{Dallaert2000sfm,Schonberger2016colmap}.
In Laue tomography, two-dimensional images are taken after rotating the sample in 3D.
While those problems are different, they share many similarities, which suggests that the EM approach to be promising for the Laue reconstruction.

Registration of point clouds is also a classic problem in CV tackled by multiple classic approaches \cite{Besl1992icp}.
A major difficulty in this problem is estimation of a rotation matrix without correspondence, which has multiple minima.
The Go-ICP algorithm finds a global minimum using a branch-and-bound method.
Robust point matching \cite{Gold1998pointmatching} proposes to use a soft assignment scheme, which can be viewed as an optimal transport problem solved by the Sinkhorn algorithm \cite{Sinkhorn1964sinkhorn}.
It uses a deterministic annealing schedule to avoid getting stuck in local minima.
As detected and modeled spots can be viewed as point clouds registration problem, this method may also have relevance for Laue diffraction.

As the number of grains is not known a-priori, Laue reconstruction has also similarities to sparse feature selection problems in the context of optimal transport.
Recently, sparse optimal transport methods have been proposed \cite{blondel2018smooth, Schmitzer2019stabilized,liu2022sparsity}.
However, the sparse OT problem is not convex and most of the methods use gradient-based optimization, which can limit the scalability of this approach.
Other methods that exploit submodular properties of some OT problems \cite{AlvarezMelis2018StructuredOT,Gurumoorthy2021spot}, which enable fast approximate optimization with convergence guarantees.

\section{Laue analysis}
\label{sec:problem}

In this section we provide a pedagogical introduction to crystallogaphic lattices, Laue diffraction, and tomography experiments.

\subsection{Crystallographic planes}

In a single crystal, the atoms are arranged in a specific pattern, described by the crystallographic point group.
The lattice of the crystal creates a set of Miller indices, determined by integer triples $[h,k,l]$.
They denote a family of parallel lattice planes, sharing the distance between atoms $a$ and normal vector $u$. 
Given three lattice vectors $a_1,a_2,a_3$ that define a unit cell, $[h,k,l]$ denotes planes that intercept the three points $[h a_1, k a_2, l a_3]$ or their multiples (for zero indices, the intercept is at infinity and the planes do not intersect the axis).
For cubic crystals with lattice constant $a$, the spacing $d$ between adjacent [$h,k,l$] lattice planes is $d = a  / \sqrt{h^2 + k^2 + l^2}$.
In this work we will use the definition of Miller indices as the inverse intercepts along the lattice vectors.
It is also common to use Miller indices defined as points in the reciprocal lattice, which we do not use here.
See \cite{Hammond2015basics} for a comprehensive introduction to crystallography.

\subsection{Laue diffraction}

\begin{figure}
\centering
\includegraphics[width=1\linewidth]{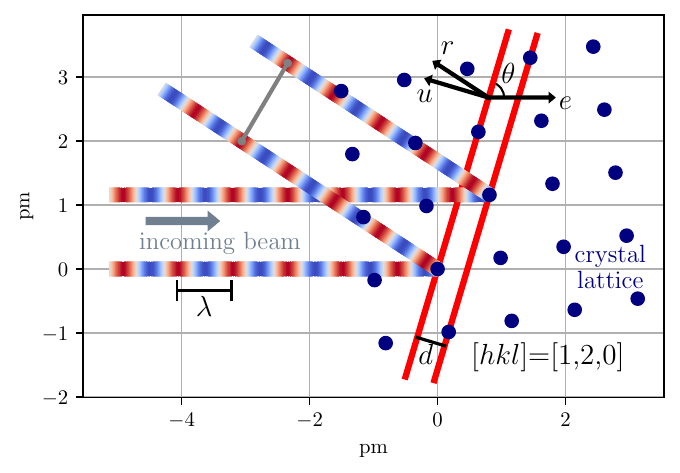}
\caption{Bragg back-scatter diffraction from a crystal lattice with cubic symmetry (pure iron). 
Showing a cross-section through the z-dimension.
Example Miller planes with normal vector $u$ and indices $[h,k,l]{=}[1,2,0]$ are shown with the red lines.
The incoming beam with wavelength $\lambda$ is shown with the colored line, with color corresponding to the value of the light sine wave.
The diffracted ray $r$ displays constructive interference with aligned phases, shown with the gray line.
}
\label{fig:bragg}
\end{figure}

Laue-Bragg interference describes the scattering of waves from a crystal lattice. 
The constructive interference of diffracted waves occurs for specific combinations of the incidence angle $\theta$ between the plane and the incoming beam, the wavelength $\lambda$, and the distance $d$ between atoms.
Bragg's law states that constructive interference occurs when
\begin{equation}
\label{eqn:bragg}
  n \lambda = 2 d \sin \theta,
\end{equation}
where the integer $n$ is the diffraction order.
For an incoming beam with unit direction $e$, the wave diffracted by plane with normal unit vector $u$ will have the unit direction $r$, which can be calculated using the Householder equation, and the wavelength $\lambda$ at which the scattering occurs
\begin{align}
r &= \pm (\mathbb{I} - 2 u u\T) e, \label{eqn:ray_r}\\  
\lambda &= 2d u\T  e \label{eqn:wavelength_lambda},
\end{align}
with ``-'' for transmission (forward-scattering) and ``+'' for reflection (back-scattering).
The intensity of the ray $r$ scales with wavelength $\lambda^4$, the incoming beam intensity at $\lambda$ and the volume of the crystal.
See Figure~\ref{fig:bragg} for an illustration of Bragg diffraction for a single plane with $[h,k,l]$=[1, 2, 0].

Let's consider a laboratory coordinate system with the crystal at its origin.
While each crystallogaphic plane in 3D has some small offset with respect to the center of the lattice, we will will use a common position $x \in \mathbb{R}^3$ in the laboratory coordinate system for all planes belonging to the crystal, which will correspond to its center.
We use the incoming beam with direction \mbox{$e{=}(1,0,0)$}, and an example detector centered at $\gamma$, with surface normal vector $\nu$.
For a particular $[h,k,l]$ Miller planes defining the diffracted direction $r$, the ray will create a Bragg peak (also called a \emph{spot}) on the detector, at the position
\begin{align}
\label{eqn:spot_position}
s &= x + \frac{\nu\T(\gamma -x)}{\nu\T r} r,
\end{align}
where $(\nu\T(\gamma -x))/(\nu\T r)$ is the distance from the center of the grain to the (3D) position of the spot in the detector.

\subsection{Tomography experiments}
\label{sec:tomo_experiments}

Laue tomography experiments aim to study the internal structure of polycrystalline samples.
A polycrystalline sample is composed of multiple single crystals, or \emph{grains}.
Each grain is described by its position $x$ inside the sample and orientation matrix $R{\in}SO(3)$ with respect to the reference $R^{\mathrm{ref}}{=}\mathbbm{I}$.
We consider a sample to consist of a set of grains $\{ \mathcal{G}_n  \}_{n\in N}$,
each characterized by its position and orientation matrix $\mathcal{G}_n{=}(R_n, x_n)$.
Each grain will contribute a set of Miller planes, which we will consider to be located at the grain's center $x$, as described above.
For a given experiment, we can limit the number of considered Miller planes, as higher order planes will not satisfy Bragg's condition because the associated wavelength is too short or too long. 
We will use a set $\{ \mathcal{M}_m \}_{m \in M}$ of plausible Miller planes, each with a corresponding unit direction $w$ and atom spacing $d$, creating a couple $\mathcal{M}_m{=} (w_m, d_m)$.
This plausible set is provided as an input to the analysis.

\begin{figure}
\centering
\includegraphics[width=1\linewidth, trim={3.5cm 2cm 3.9cm 3.3cm},clip]{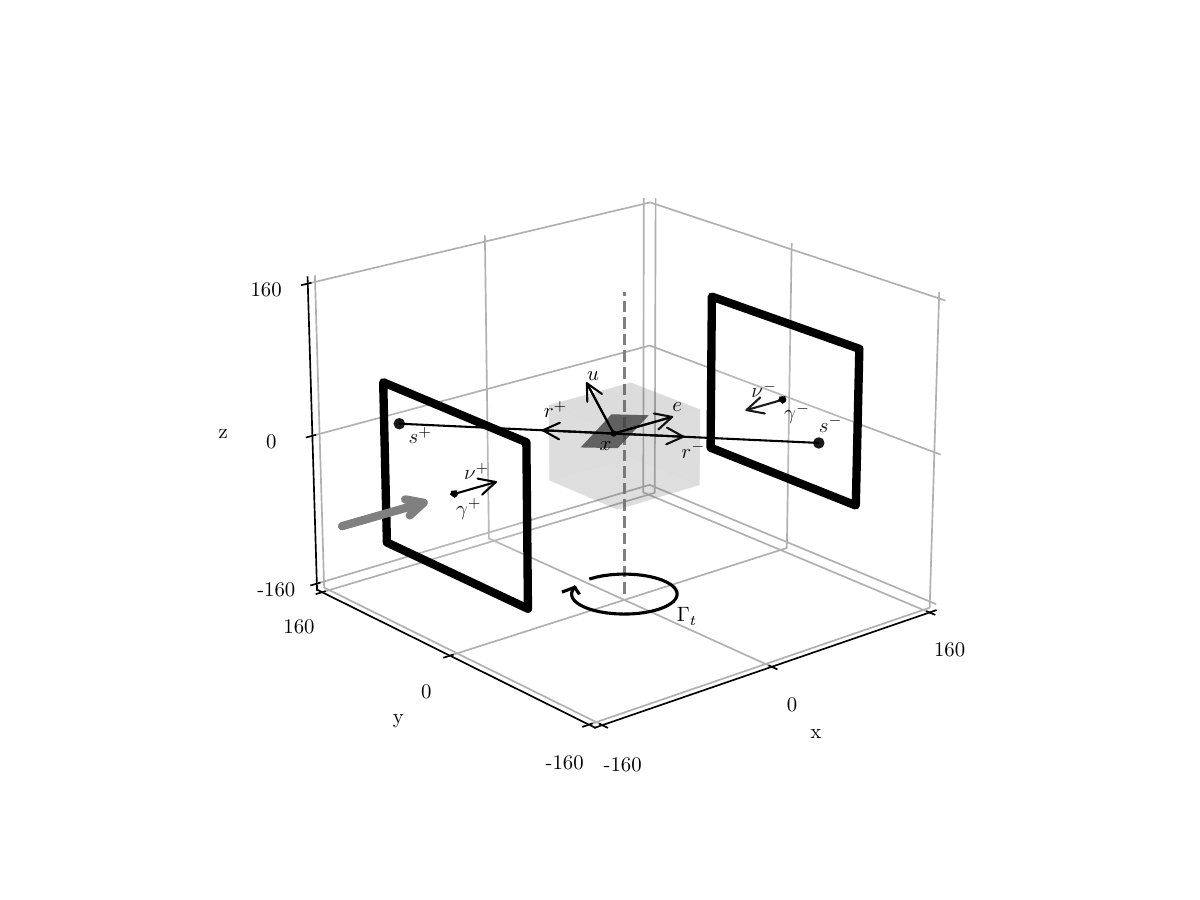}
\caption{A scene of a tomographic diffraction experiment with two detectors configured for recording backscatter and transmission spots.
The thick gray arrow corresponds to the incoming beam.
The beam direction is set by vector $e$.
The light gray cube is the studied sample.
The dark gray plane is an example Miller planes in a grain.
The plane is oriented inside the grain with normal vector $u$.
The constructive interference will occur at directions $r^{-}$ (transmission) and $r^{+}$ (backscatter) and create spots $s^{-}$ and $s^{+}$, respectively, on the detector screens.
For clarity, the detector size used here is smaller than in a typical experiment.}
\label{fig:schematic}
\end{figure}

The positions and orientations of the crystal are found by recording locations of spots created by diffracted rays on the detector screen after illuminating the sample by a wide-spectrum beam. 
We will consider the beam to have the wavelength $\lambda$ in range $\lambda{\in}[\lambda^{\mathrm{min}}, \lambda^{\mathrm{max}}]$.
During a tomographic experiment, the sample is illuminated multiple times after being rotated. 
Let $\mathcal{T}{=}\{ \Gamma_t \}_{t\in T}$ be a set of rotation steps, where $\Gamma_t{\in}SO(3)$ is the rotation matrix for step~$t$.
The formulations that we will develop are applicable to any rotation, but the experimental setup that we consider allows in practice only rotations along the (0,0,1) axis.
Figure~\ref{fig:schematic} shows the scene in question, with two detectors, recording back-scatter or forward-scatter rays.
The thick grey arrow shows the incoming beam, the thin dashed line corresponds to the axis of rotation for tomographic projections.
A single plane shown in dark grey, while the light gray shows the studied sample in which the plane is located.
In general, more detectors can be included in the experiment. 
In the interest of clarity, we will from now on consider only the backscatter detector and rays.
An equivalent analysis can be easily performed for the forward-scattering mode.

At a given sample rotation $\Gamma_t$ and for grain orientation $R_n$, the vector $u_{mnt}$ normal to the rotated Miller plane $w_m$ will be
\begin{equation}
u_{mnt}=\Gamma_t R_n w_m,
\end{equation}
the model ray is emitted in the direction 
\begin{equation}\label{eqn:ray_mnt}
    r_{mnt}=  (\mathbb{I} - 2 \Gamma_{t} R_{n} w_{m} w_{m}\T R_{n}\T\Gamma_{t}\T) \, e\, ,
\end{equation}
creates a spot on the detector plane at position $s_{mnt}$ as follows
\begin{align}\label{eqn:s_mnt}
    s_{mnt} &=\Gamma_t \,x_n+ \frac{\nu\T(\gamma - \Gamma_{t} x_{n})}{\nu\T r_{mnt}} r_{mnt},
\end{align}
and at wavelength $\lambda_{mnt}$ with
\begin{equation}
    \lambda_{mnt} = 2 d_m (\Gamma_{t} R_{n} w_{m})\T e.
\end{equation}    

\subsubsection{Detected spots}

We assume that the center $\gamma$ of a square backscatter detector lies on the x-axis of the laboratory coordinate system, with size $l$ mm per side.
We will assume that $\gamma$ and $l$ are known, as they can be characterized accurately with a proper calibration process \cite{samothrakitis2023calibr}.
The detection of spots and the measurement of their positions is performed by an external algorithm and is considered as input to this analysis.
Next to real Bragg peaks, the set of detected spots will also contain spurious detections, which we discuss in Section~\ref{sec:outliers}.
An experiment will yield a collection of \textbf{detected spots} $\{ p_i \}_{i \in I}$, in all experimental images obtained for different rotations of the sample. 
We will denote by $t_i$ the rotation step which produced the $i$th spot.

We assume a ray noise model, where a noisy unit ray direction follows the von Mises-Fisher distribution in 3 dimensions \cite{Fisher1953sphere}, with mean $r$ and concentration $\kappa$
\begin{equation*}\label{eqn:ray_mnt_noisy}
    \mathbf{vMF}_3( y \ | \ r, \kappa ) =  \frac{\kappa}{4\pi \sinh \kappa } \exp \left(\kappa \cdot r\T y \right).
\end{equation*}
Note that samples from this distribution can be obtained by drawing from a 3D isotropic multivariate normal distribution with covariance $ C{=}\sigma^2 \mathbbm{I}$, where $\sigma^2{=}1/\kappa$, and then conditioning $\| x  \|=1$.
Therefore the noisy ray $\hat r_i$ is 
\begin{equation}
  \tilde r_i \sim \mathbf{vMF}(r_i,\sigma^{-2}),
\end{equation} 
and the detected spots will be offset from their true positions $s_i$ following the noisy ray $\tilde r_i$
\begin{align}
\label{eqn:spot_noisy}
    s_i &=\Gamma_{t_i} x_{n_i} + \frac{\nu\T(\gamma - \Gamma_{t_i} x_{n_i})}{\nu\T \tilde r_i} \tilde r_i.
\end{align}

\subsubsection{Model spots}

In our method, we aim to fit a dataset of detected spots with a set of \textbf{model spots} stemming from some candidate model sample in our experimental setup.
Not all of the model spots $s_{mnt}$ can be considered in the problem, for three reasons.
Firstly, the model spot can lie outside the detector area.
Secondly, the brightness of the Bragg peak will be proportional to the intensity of the beam at the corresponding wavelength $\lambda$; if the intensity is close to zero, then the spot will not be detected.
Finally, the intensity of the spot will be proportional to the volume of the grain; small grains will yield low-intensity spots that will not be detectable.

We therefore use a sequence of model spots $\mathcal{S}{=}(s_j)_{j \in J}$, that contains only those spots that simultaneously lie inside the beam wavelength range and the detector screen area
\begin{align}
\begin{aligned}
\label{eqn:model_spot_cut}
  \mathcal{S} = \{ s_{mnt} : \ & (m,n,t) \in M \times N \times T  \\ & \land  \ \lambda_{mnt} \in [\lambda_{\mathrm{min}}, \lambda_{\mathrm{max}}] \  \land  \ \| s_{mnt} - \gamma \|_{
  \infty} < l/2  \}.
\end{aligned}
\end{align}
We will use a sequence $(r_j)_{j\in J}$ to denote rays corresponding to spots in $\mathcal{S}$.
For clarity, we will use a sequence $(t_j)_{j\in J}$ of indices $t_j{\in}T$ to label the step at which spot $j$ was generated, itself labeled by index set $J_t$.
Equivalently, let sequence $(n_j)_{j \in J}$ of indices $n_j{\in}N$ denote the index of the grain at which spot $j$ was created, itself labeled by index set $J_n$.
Thus, $|J_n|$ and $|J_t|$, respectively, will be the number of model spots generated by grain $n$ or step $t$, meeting the criteria in Equation~\ref{eqn:model_spot_cut}.
Finally, let $m_j$ denote the index of the Miller plane which refracted the ray producing spot $j$ and $J_m$ the collection of spot indices associated with a particular Miller plane $m$.

\section{The likelihood}
\label{sec:likelihood}

In this section, we propose a probabilistic model for the collection of observed spots detected given a collection of model spots produced by a set of grains $N$ producing each a refraction for a collection of Miller planes $M$, with latent variables encoding which detected spot matches which model spot.
We assume the experimental setup defined above with detector described by ($\gamma,\nu,l$), with sample rotation step in $T$ yielding detected spots indexed by $I$.
First, we introduce the problem without outlier spots, followed by introduction of the outliers. 

Let's consider a guess grain with position $x_{n}$ that generates the ray $j$. 
For a detected spot $s_i$ is matched with model ray $j$, we define a ray estimate $\hat r_{ij}$ as
\begin{align}
  \hat r_{ij} = \frac{s_i - \Gamma_{t_j} x_{n_j}}{\| s_i - \Gamma_{t_j} x_{n_j} \|}.
\end{align}
This quantity will be useful in further calculations.

\subsection{Spot-to-spot assignment}

We assume that there are no outliers among detected spots and no missing model spots (corresponding to unmodeled Miller planes), so that there is a one-to-one correspondence between model and observed spots.
To match the detected spots with model rays, we introduce the assignment variables $Z_{ij}{\in}\{0,1\}: i{\in}I,\, j{\in}J$, with $Z_{ij}{=}1$ if the detected spot $s_{i}$ is associated to \mbox{model ray $r_j$}, and $Z_{ij}{=}0$ otherwise.
Only matches between spots occurring at the same sample rotation $t$ are allowed, otherwise $Z_{ij}{=}0$.
Each $i$ is matched exactly to one $j$, so the joint log-likelihood \mbox{$\ell:=\sum_{i=1}^n \log p (s_i\mid Z_{i\cdot},s)$} for all observed spots $s_i$ given $Z_{i\cdot}:=(Z_{ij})_{j \in J}$ and $r_j$ as 
\begin{align}
\label{eqn:likelihood}
   \ell= \frac{1}{\sigma^2} \sum_{i \in I} \sum_{j \in J}  Z_{ij} \cdot \hat r_{ij}\T r_j - c.  
\end{align}
where $c$ is the log normalizing constant.
At step $t$, the number of models spots and detected peaks is $|J_t|$ and $|I_t|$, respectively.
Note that if we assume a uniform prior probability for the assignment of $i$ to all $j$s such that $t_i=t_j,$ and with $Z:=(Z_{ij})_{i \in I, j\in J},$ we have
\begin{align} 
p(Z)=\prod_{t}\prod_{(i,j):t_i=t_j=t} |J_t|^{-Z_{ij}}=|J_t|^{-|I_t|},
\end{align}
the joint log-likelihood only differs from $\ell$ by $-\sum_t |I_t| \log |J_t|,$ which just changes the normalizing constant~$c$.
For clarity, we denote the spot assignment likelihood matrix as $L\in\mathbb{R}^{|I|\times|J|}$.
It contains the cost of assigning observed spot $i$ to model spot $j$
\begin{equation}
L_{ij} = \frac{1}{\sigma^2} \hat r_{ij} \T r_j \quad  i \in I, \ j \in J.
\end{equation}
Since the assignment of peaks to models spots is a-priori unknown, treating the variables $Z_{ij}$ as latent variables and maximizing the marginal log-likelihood \mbox{$\sum_i \log p (s_i \mid (R_n, x_n)_{n\in N})$} can be done with a classical EM-algorithm, which maximizes the evidence lower bound (ELBO)
\begin{equation}
\mathcal{L}(Q;(R_n,x_n)_n)= \sum_{i\in I} \sum_{j \in J}  Q_{ij} L_{ij} - const.,
\end{equation}
where $Q \in \mathbb{R}_+^{|I| \times |J|}$ is the responsibility matrix with elements $Q_{ij}{=}\mathbb{E}[Z_{ij}]$.
Instead of the classical EM approach, we propose to use the Sinkhorn-EM (sEM) formulation \cite{Mena2020sinkhorn}, which was recently introduced for solving mixture model problems. 
While the classical EM method calculates the cluster memberships independently, sEM computes them using optimal transport, where the responsibilities respect the known proportions.
It it shown that sEM displays better global convergence guarantees, while optimizing the lower bound on the log-likelihood and thus maintaining the probabilistic interpretation of the analysis. 
In sEM, the likelihood is replaced with an entropic-OT likelihood
\begin{align}
\label{eqn:entropic_ot_loss}
\mathcal{L_{\text{OT}}}(Q;(R_n,x_n)_n) & = \sup_{Q \in \Pi(a,b) } \left[ \sum_{i\in I} \sum_{j \in J}  Q_{ij} L_{ij} + \mathbf{H}(Q) \right], 
\end{align}
where $\Pi(a,b)$ is a set of all transport plans that have joint distributions with marginals $a$ and $b$, respectively, and \mbox{$\mathbf{H}(Q)=-\sum_{i,j} Q_{ij} \log Q_{ij}$} is the entropy of 
$Q$\footnote{Here $Q$ refers to the transport plan, not the entropy of the distribution over permutation with mean $Q$.}.
In a discrete setting applicable here, the marginal mass distributions $a$ and $b$ have entries uniformly equal to $1$: for each detected and model spot: $a_i{=}b_j{=}1$.
This formulation requires solving an optimal transport problem.
This likelihood is consistent in the population limit with the classical likelihood approach while having a better geometrical properties and is proven to be less prone to getting stuck in local optima than classical EM.
Using this formulation enables practical solving of the large-scale Laue problem.

Note that $Z_{ij}$ and $Q_{ij}$ can only be non-zero if $t_i{=}t_j;$ in the rest of the paper, we will consider the set $\mathcal{Q}$ of matrices $Q$ such that $Q_{ij}{=}0$ for all $(i,j)$ with $t_i \neq t_j$. 

\subsection{Assignments with outlier and missing spots}
\label{sec:problem_with_outliers}

In this section we consider outlier spots, both in detected and modeled sets.
There are two reasons for outliers among spots detected in image data.
Firstly, the outliers can be just spurious detections: noise spikes or points corresponding to detection algorithm failures.
Blended overlapping spots can also have dramatically wrong position measurement, which lies outside the range allowed by the noise level.
Secondly, the sample can contain very small grains that will give rise to only few detectable spots, which can be insufficient to form a diffraction pattern that can be reliably measured by the solver.
The number of outliers in the data can be large and it is typically not known by the user.

We also consider unmatched model rays.
The area of the screen and wavelength range are limited, which leads to spots moving out and into the image as the grain is rotated.
We start with a set of initial (or \emph{prototype}) grains, giving rise to a set of model spots $\mathcal{S}$, with  spots outside the detector screen or outside the beam's wavelength range being excluded (Equation~\ref{eqn:model_spot_cut}). 
However, as the orientation and position of the grains are optimized, some spots may exit the detector/wavelength range and should be also excluded during the matching process. 
Conversely, there will be spots that were initially excluded, but would appear for the updated grain parameters.
Precise modeling of this property in the likelihood would be complicated. 
Therefore, we propose a simpler approach, were we treat these cases as unmatched model spots instead.
During the optimization we will use the same set of model spots as predicted using for the initial values of the prototypes grains with $(R^0_n,x^0_n)_{n\in N}$.
The closer the prototype model parameters are to the their true value, the fewer unmatched spots will be encountered. 
According to our empirical tests for the number of prototypes considered here, the fraction of different between the candidate models and their true counterparts is no more than 10\%.

We model the outliers using a full probabilistic description of the problem. 
We add an extra row to $Z$, with index $0$, reserved for the assignment of spurious detection to the outliers status, i.e. $Z_{i0}=1$ for outlier spots.
Symmetrically, we add an extra column, again with index $0$, to assign unmatched model spots, i.e. $Z_{0j}=1$ if a model spot does not match any detected peak. 
Note that multiple entries of the $0^{\mathrm{th}}$ row (resp.~column)
can be non-zero.
At the intersection of these rows the entry $Z_{00}$ is ignored.
To take into account explicitly the possibility of outliers into the probabilistic model, we modify the likelihood as follows. 
We assume there is first a fixed probability $\pi_0$ for a detected spot to be an outlier, so that $p(Z_{i0}{=}1){=}\pi_0$ and \mbox{$p(Z_{ij}{=}1){=}(1-\pi_0)/|J|$}.
Given that it is an outlier, we assume that it appears uniformly over the detector so that $p(s_i|Z_{i0}=1)=\frac{1}{S}$ where $S$ is the surface of the detector, if it is not an outlier, we use the same model as before.
The joint likelihood of a detected spot position $s_i$ and it assignment variable $Z_{i\cdot}$ can then be written as
\begin{align}
\begin{aligned}
    p(s_{i}, Z_{i \cdot}) = \left( {\textstyle \frac{\pi_0}{S}} \right)^{Z_{i0}} \prod_{j=1}^{J} \left( {\textstyle \frac{1}{c}\exp L_{ij}}  \right)^{Z_{ij}} 
  \left( {\textstyle \frac{1-\pi_0}{J}} \right)^{Z_{ij}}. 
\end{aligned}
\end{align}

\noindent
Each term of the log-likelihood can be written as 
\begin{align*}
\begin{aligned}
  \ell(s_{i}, Z_{ij}) = Z_{i0}\log {\textstyle \frac{\pi_0}{S}} &+ (1-Z_{i0})\log \textstyle{ \frac{1-\pi_0}{J c}} +
  L_{ij} Z_{ij}.
\end{aligned}
\end{align*}

\noindent
Given that our likelihood is constructed per detected spot, and that we do not model their number, the probability that a model spot is unmatched is modeled implicitly, given the total number of model spots. 
We do not explicitly model the probability of unmatched model spots in our formulation of the problem, and set it to likelihood value of the von Mises-Fisher distribution with concentration $\kappa{=}\sigma^{-2}$ that corresponds to the null angle difference. 
Note that, up to additive constants, the term for a pair $(i,j)$ with $i,j \neq 0$ in the log-likelihood remains as before.
We define the extended likelihood matrix
\begin{align}
\begin{aligned} 
\label{eqn:update_L}
    L_{0j} &= \sigma^{-2} - \log(4 \pi \sigma^2 \sinh (\sigma^{-2})) \\
    L_{i0} &= \log\left( \frac{\pi_0}{1-\pi_0} \cdot \frac{|J| c}{S} \right) =: L^{\rm{out}}\\
    L_{ij} &= \frac{1}{\sigma^2}  \hat r_{ij}\T r_j \quad \forall i{\in}I, j{\in}J.
\end{aligned}
\end{align}
As in practice it is difficult to estimate $\pi_0$, in our implementation we use a fixed outlier likelihood $L^{\rm{out}}$, as defined in Equation~\ref{eqn:update_L}.
We set $L_{00}{=}L^{\rm{out}}$; the corresponding assignment $Q_{00}$ will serve as a sink for unused outlier capacity.
The complete log-likelihood of the data and associated assignments take the form \mbox{$\langle L, Z \rangle-const. + \log \left( {\textstyle \frac{1-\pi_0}{J c}} \right),$} where now both $L$ and $Z$ have an extra row and column.
With constants dropped, the Sinkhorn EM objective is
\begin{equation}
\begin{aligned}\label{eqn:elbo}
\mathcal{L_{\text{OT}}}(Q;(R_n,x_n)_{n \in N})= \langle L, Q \rangle + \epsilon\mathbf{H}(Q)
\end{aligned}
\end{equation}
where $Q \in \Pi(a, b)$ is the entropic transport plan, $a$ and $b$ are marginal constraints (see Section~\ref{sec:estep_sinkhorn}), and $\epsilon$ is the entropic regularization parameter, which is set to $\epsilon{=}1$ in standard sEM (also see Section~\ref{sec:annealing}).
In literature, there are other optimal transport problem variants that allow for outlier treatment, such as unbalanced OT \cite{Chizat2018scaling} and partial OT \cite{Figalli2010partial}.
We compare the performance of our proposed outlier model with these other variants.
See Appendix~\ref{app:other_ot_methods} for details.

\subsection{E-step with Sinkhorn updates}
\label{sec:estep_sinkhorn}

The E-step for finding the assignment $Q$ now requires to solve
\begin{equation}
\begin{aligned}\label{eqn:estep_obj}
\displaystyle{\max_{Q}} \quad &\left\langle L, Q \right\rangle + \epsilon \mathbf{H}(Q) \\[0.5em]
s.t.                          \quad &  Q \geq 0, \quad Q \in \mathcal{Q}, \quad Q\mathbbm{1} = a, \quad Q\T\mathbbm{1} = b,
\end{aligned}
\end{equation}
which follows from Equation~\ref{eqn:elbo}, and $\epsilon$ is the entropic regularization parameter (also see Section~\ref{sec:annealing}).
The maximum number of allowed outliers can be encoded in vectors $a$ and $b$, with $T_a$ and $T_b$ corresponding to the maximum number of spurious detections and unmatched model spots, respectively.
We set $a_i{=}b_j{=}1, \ i,j{>}1$ for matched spots, so the weight of the outliers is $a_0{=}T_{a}$ and $b_0{=}T_{b}$.  
In order for the double-stochastic constraint in Equation~\ref{eqn:elbo} to be satisfied, we use $T_b$ as the user-controlled variable and set the $T_{a} = T_{b} - b\T \mathbbm{1} + a\T \mathbbm{1}$.
The transport plan matrix is updated using the Sinkhorn subroutine 
$Q \leftarrow \textbf{BalancedOT}[L, \epsilon, \zeta^{\rm{tol}}]$
shown in Algorithm~\ref{alg:sinkhorn_balanced}, described in Appendix~\ref{alg:sinkhorn_balanced}.
The convergence of this algorithm is controlled by user-specified parameter $\zeta^{\rm{tol}}$.

\subsection{The M-step updates}
\label{sec:fitting_rot_pos}

The M-step finds the maximum likelihood orientations and positions for all grains $(R_n, x_n)_{n\in N}$ by solving a separate problem for each grain
\begin{equation}\label{eqn:mstep_obj}
\begin{aligned}
\displaystyle{\forall n\in N : \max_{x_n, R_n : R_n\T R_n = \mathbbm{I}}} \quad &\left\langle L, Q \right\rangle
\end{aligned}
\end{equation}
We solve this problem by coordinate descent in $x$ and $R$.
We propose to find grain position $x$ using Newton iterations. 
We calculate the update at step $k$ as
\begin{align}
  \Delta x^{k+1} &= \left( H(x^{k}) + \iota \mathbbm{I} \right)^{-1} g(x^{k}),
\end{align}
where $H(x^{k})$ and $g(x^{k})$ are gradient and Hessian matrices evaluated at the current position $x^{k}$, and $\iota$ is a very small constant added for numerical stability. 
We show the calculation of these matrices in Appendix~\ref{app:mstep}.
The update at iteration $k$ is
\begin{align}
x^{k+1} &\leftarrow x^{k} - \Delta x^{k}.
\end{align}
The iterations are repeated until convergence threshold $\zeta^{x}$ is met: $\| x^{k+1} - x^{k}\| < \zeta^x$.
Subroutine $\textbf{LauePosNewton}$ (Algorithm~\ref{alg:posnewton}, described in Appendix~\ref{app:mstep}), shows the details of this process.

Finding the M-step update for the orientation matrix is more complicated.
In Appendix~\ref{app:mstep} we derive a Majorization-Minimization algorithm for finding the orientation matrix $R$ maximizing the loss $\mathcal{L_{\text{OT}}}$.
This sequence starts from an initial rotation guess $R^0$.
We show that the maximization of the objective is equivalent to the maximization of a concave quadratic, for which a local maximum can be obtained by solving a sequence of linear optimization problems over rotation matrices that each take the form of a \emph{Wahba's problem} (an alternative formulation of the \emph{Orthogonal Procrustes problem}, see Appendix~\ref{app:wahba} for details).
First, we calculate the matrix $M_m^0$ using the current values of grain positions $x_n$ and assignments $Q$.
We then shift it so that it becomes PSD by adding a constant diagonal $\iota \mathbbm{I}$
\begin{align}\label{eqn:majmin_step}
M^0_m &:= -\sum_{j \in J_n \cap J_m } \Gamma_{t_j}^\top \Big [ \sum_{i\in I} Q_{ij} \, (\hat r_{ij}e^\top + e \hat r_{ij}^\top ) \Big ] \Gamma_{t_j}, \\
M_m &:= M_m^0 + \iota \mathbbm{I}
\end{align} 
where $w_m$ is the direction of ray $m$ and $M_m$, and $\iota$ is the absolute value of the most negative eigenvalue of $M_m^0$ or 0 if $M_m^0$ does not have negative eigenvalues.
The update at iteration $k$ is 
\begin{align}
  R^{k+1} &\leftarrow \arg \min_{R: R^\top R=\mathbbm{I}} \sum_{m} \|R\,  w_m- M_m R^k w_m\|^2.
\end{align}
The iterations terminate when the convergence threshold $\zeta^{R}$ is met: $\| R^{k+1} - R^{k}\|_F < \zeta^R$.
Subroutine $\textbf{LaueRotMM}$ (Algorithm~\ref{alg:rotmajmin}, described in Appendix~\ref{app:mstep}) presents this method in detail.

\section{The \ours\ solver}

\begin{figure*}
\centering
\includegraphics[width=1\linewidth]{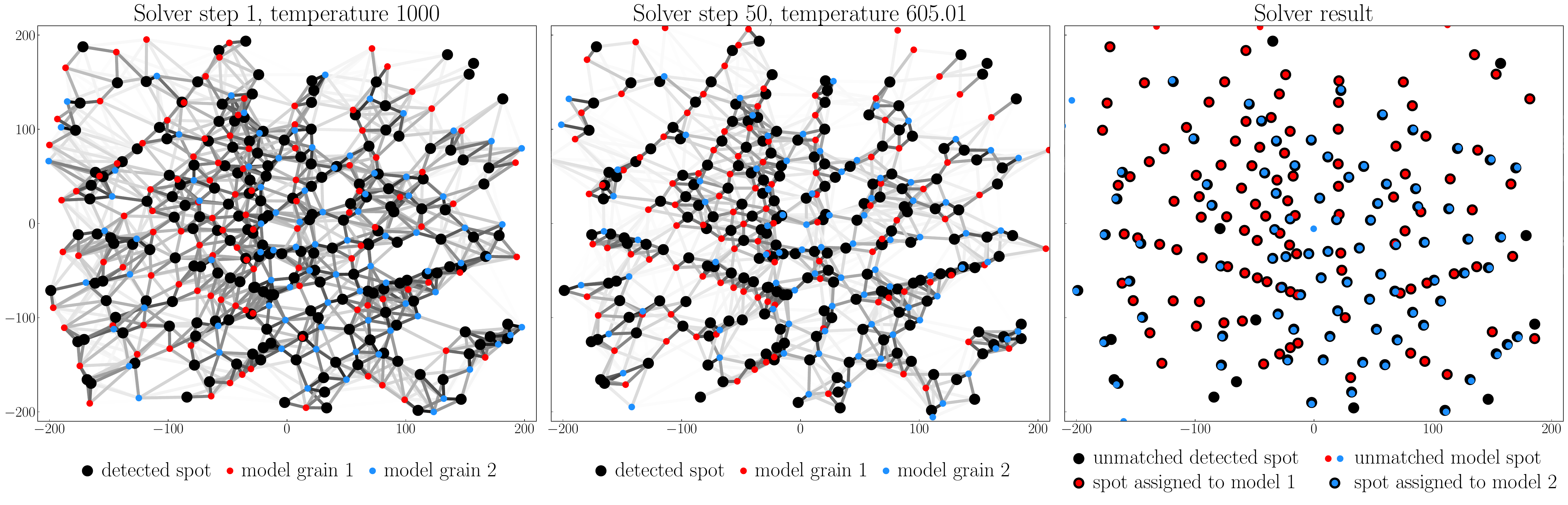}
\caption{
Illustration of three selected steps of the \ours\ solver on a toy problem with a simulated ruby-like sample with two grains and noise-free detected spots, for a single projection in the backscatter detector.
The left panels shows the initialization of two grains with red and blue model spots, and strength of the initial assignment weights in grey lines.
The darker the color of the line, the stronger the assignment weight. 
The algorithm starts with a high temperature, which enables assignments between model and detected spots that can be far. 
As the optimization progresses, the temperature is lowered, as shown in the middle panel.
Finally, the algorithm assigns the model and detected spots to the grains.
Note that some of the detected spots were not matched; those will be considered as outliers.
Conversely, some model spots are unmatched.
}
\label{fig:steps}
\end{figure*}

To efficiently solve the stated problem at scale, we propose the \ours\ formulation (Laue indexing with Optimal Transport).
We introduce key elements of this method below.

The \ours\ method uses iterative updates for all variables.
This enables very efficient implementation of coordinate ascent steps on GPUs, as all of these operations require only simple tensorized algebra.

As the number of local maxima in this problem can be large, techniques for global optimization can be employed. 
In this work we use a deterministic annealing scheme to control the ``temperature'' of the problem \cite{Gold1998pointmatching}\cite{Rangarajan1997softassignprocrustes}.
It enables more efficient search for local maxima.
During the final steps of the algorithm the temperature control is switched off, which corresponds to finding the local maximum likelihood in Equation~\ref{eqn:elbo}.
This process is described in Section~\ref{sec:annealing}.

In a typical Laue analysis, the number of spots in the problem can be large, up to several millions. 
Additionally, only matches of spots at the same sample rotation $t$ are allowed. 
To enable good performance of the method at scale, we work with a sparse representation of the assignment matrix (see Section~\ref{sec:neighbors}).

The multi-grain fitting step is initialized using a set of prototype grains, which together constitute the final sample.
Therefore, finding a good set of prototype grains is a critical step.
We create the set of prototype grains using two pre-processing steps: 
(i) single-grain fitting, where we find local minima for samples consisting only of a single grain, starting from multiple initialization points in rotation and position space, and 
(ii) prototype selection, where we select a small number of prototypes out of all candidates, that form the final sample.
We describe the pre-processing steps in Section~\ref{sec:preprocessing}.

\subsection{Deterministic annealing}
\label{sec:annealing}

The problem in Equation~\ref{eqn:elbo} is non-convex and has multiple local minima. 
This is a common problem for EM algorithms.
Deterministic annealing variant of EM has been proposed to address this \cite{Ueda1994emannealing}.
In this variant, the likelihood function of data $x$ and latent $y$ given parameters $\Theta$ is replaced by another function
$p(x,y \ | \ \Theta) \rightarrow p(x,y \ | \ \Theta)^\beta$, where $\epsilon{=}1/\beta$ is the ``temperature''.
The temperature starts at a value $\epsilon{\gg}1$ and is gradually lowered during the EM iterations.
At $\epsilon{=}\beta{=}1$, the original EM formulation is recovered.
Gradual temperature decrease helps with avoiding local minima.
Similar scheme that additionally includes outlier variables was used in \cite{FruhwirthWaltenberger2016annealing}.
In the Laue problem, using temperature $\epsilon{>}1$ is analogous to increasing entropic regularization in optimal transport.
Starting with large $\epsilon$ allows for many rays to explain a given detected spot.
Decreasing $\epsilon$ allows to to lift gradually the assignment ambiguities.
The initial temperature $\epsilon^{\rm{init}}$ is typically chosen to be $\epsilon^{\rm{init}}{>}100$.

\subsection{Working with sparse assignment matrices}
\label{sec:neighbors}

Considering only $\rho$ nearest neighbors in the assignment scheme, which is equivalent to setting null assignment weight in the full $Q$ matrix, has the strong advantage to decrease the complexity of Sinkhorn updates from $J^2$ to $\rho J.$
In the Sinkhorn algorithm, this will be a good approximation as long as $\exp(L/\epsilon)$ is close to zero for all points outside the neighbor set. 
We find this approximation to be stable enough for our implementation, although other approaches could be employed here \cite{Schmitzer2015sparseot,Schmitzer2019stabilized}.

\subsection{The algorithm}
\label{sec:algo}

The \ours\ method is described in Algorithm~\ref{alg:multigrain} (Appendix~\ref{app:algos}).
Next to the detected spots and algorithm control parameters, the routine takes as input a set of prototype grains $(R_n, x_n)_{n \in N}$, which is calculated in the pre-processing steps.
A simple visual demonstration of the algorithm is shown in Figure~\ref{fig:steps}.
It shows two candidate grains (blue and red spots) being fit to a set of detected spots (black), at the initial optimization step (left panel), step 50 (middle panel), and the final assignment (right panel).
The gray lines show the strength of the assignment $Q$.
Note the outliers (black circles) and unmatched model spots (colored circles) in the final assignment.

\section{Pre-processing}
\label{sec:preprocessing}

We perform two preprocessing steps: 
\begin{enumerate}
\item[(i)] single-grain fitting, where we fit multiple candidate single grains from different starting points in rotation and position space, 
\item[(ii)] selection of prototypes from the list of optimized candidate grains using the \emph{Selection of Prototypes with Optimal Transport} (SPOT) framework \cite{Gurumoorthy2021spot}.
\end{enumerate}
The goal of the first step is to find single grains that fit the data well and can be included in the final grain set.
The second step selects the final set of prototype grains that will be included in the sample and considered for multi-grain optimization.

\subsection{Single-grain fitting}
\label{sec:single_grain}

The first pre-processing step is a single grain search over the space of its orientations and positions.
The necessity for this step stems from the fact that, even for a single grain, the problem in Equation~\ref{eqn:elbo} has multiple local minima in the $x$ and $R$ variables.

Generally, the problem of finding orientation matrices to match point sets is multi-modal for data without correspondences \cite{Besl1992icp,Rangarajan1997softassignprocrustes} and is typically solved by running convex solvers multiple times from different starting points in $SO(3)$.
We note that, for many other computer vision problems, a good set of starting points can be pre-computed using simple alignment of principal components axes in source and target sets.
Diffraction patterns, however, do not have a specific ``shape'', which prohibits this approach.
As a pre-processing step to find a large number of candidate grains, we use single-grain fitting runs, initiated a different starting points in the six-dimensional $(R,x)$ space (see Section~\ref{sec:single_grain}).
We discretize the search space using a Sobol sequence grid, where the grain orientation is represented using the \emph{modified Rodrigues parameters}~\cite{Terzakis2018mrp}.
The search over $R$ is restricted according to the crystalographic symmetries of the lattice \cite{He2007lattice}, while the search over $x$ is limited to points within the bounds of sample size.
These limits are set by the user depending of the type of lattice corresponding to the composition of the sample, as well as its physical size.

From each initialization point, we run a single grain fitting using a single-grain variant of the problem, described in Appendix~\ref{app:single_grain_fitting}.
This formulation will treat most of the detected spots as outliers.
The number of runs is variable, depending on the expected number of grains in the sample.
For 10 and 1000 grains, we run 5k and 500k initizalizations, respectively.
Once the optimization is finished, we store the converged grain orientations and positions.
Subroutine \textbf{SingleLaueOT} (Algorithm~\ref{alg:single}, described in Appendix~\ref{app:single_grain_fitting}) is used for this task.

\subsection{Selection of Prototypes with Optimal Transport}

The goal of this step is to select the grain candidates that will compose the final sample. 
As no further grains will be added to the sample later, this is a critical part of the algorithm.
Selecting too many or too few grain candidates can lead to errors in the multi-grain fitting solutions.
We start with removing badly-fitted models by selecting grains that have the fraction of unmatched model rays grater than $f^{\rm{out}}$.
A model ray is considered unmatched if its largest weight in $Q$ is assigned to the outlier column.
Then to create the final sample, we perform prototype selection in the framework \emph{Selection of Prototypes using Optimal Transport} (\spot) \cite{Gurumoorthy2021spot}.

The \textsc{Spot} framework casts the problem of selection prototypes as a problem of learning a sparse source probability distribution supported on $k$ elements that has the minimum OT distance from the target distribution. 
It poses the prototype selection problem as a maximization version of the OT problem, where the source distribution weights are learned with a cardinality constraint.
The proposed loss has a key property of submodularity, which enables the use of efficient greedy method with deterministic approximation guarantees.
We use the \textsc{SpotGreedy} method from \cite{Gurumoorthy2021spot}, where we pose the problem as grain-to-spot OT problem of model grains to detected spots.
Algorithm~\ref{alg:prototypes}, described in Appendix~\ref{app:proto}, shows the subroutine \textbf{LaueSpotGreedy} which is an adaptation of this method for our problem.
In this algorithm, we also determine the final number of prototypes $N$: we find $N$ after which the improvement in the total cost of transport significantly decreases (see Appendix~\ref{app:proto} for details).

We note that this algorithm allows for assignment of a larger number of detected spots to the grain than its number of model spots. 
Enforcing that constraint leads to a more complicated optimization problem of support selection, recently proposed by \cite{Riaz2023partial}. 
The problem formulation presented there does not include the possibility of target sample outliers, as well as poses a difficult challenge for optimization.
As we find that the \textsc{Spot} method works well in practice, we leave more further investigation in this area to future work.

\section{Simulations and experiments}
\label{sec:experiments}

\begin{figure*}
\includegraphics[width=1\linewidth]{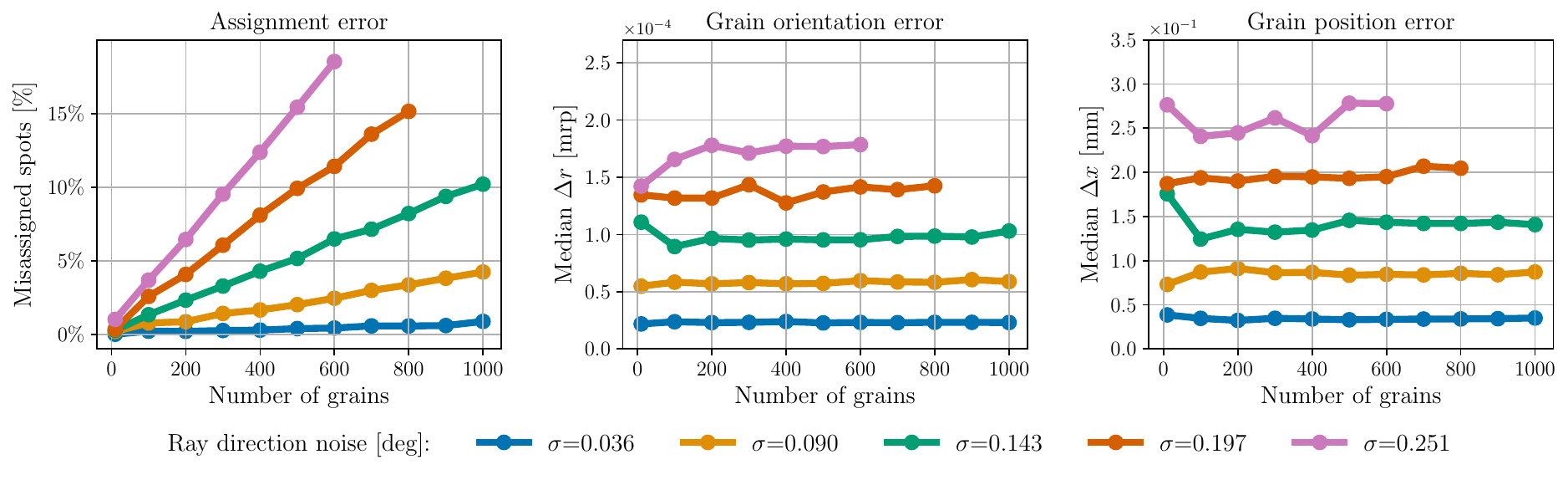}
\caption{
Quality metrics for indexing with \ours\ using simulated data with variable number of grains and noise levels, for the constant fiducial outlier fraction of 10\%.
The left panel shows the fraction of correctly assigned spots in the entire problem.
The middle and right panel show the scaled median absolute deviation of the error on the grain orientation and position, respectively. 
The number of grains was correctly identified for up to $500$ grains, and after that there was up to 1 grain missing.
The method fails for large noise levels and high grain counts.
}
\label{fig:quality_metrics}
\end{figure*}

\begin{figure*}
\centering
\includegraphics[width=1\linewidth]{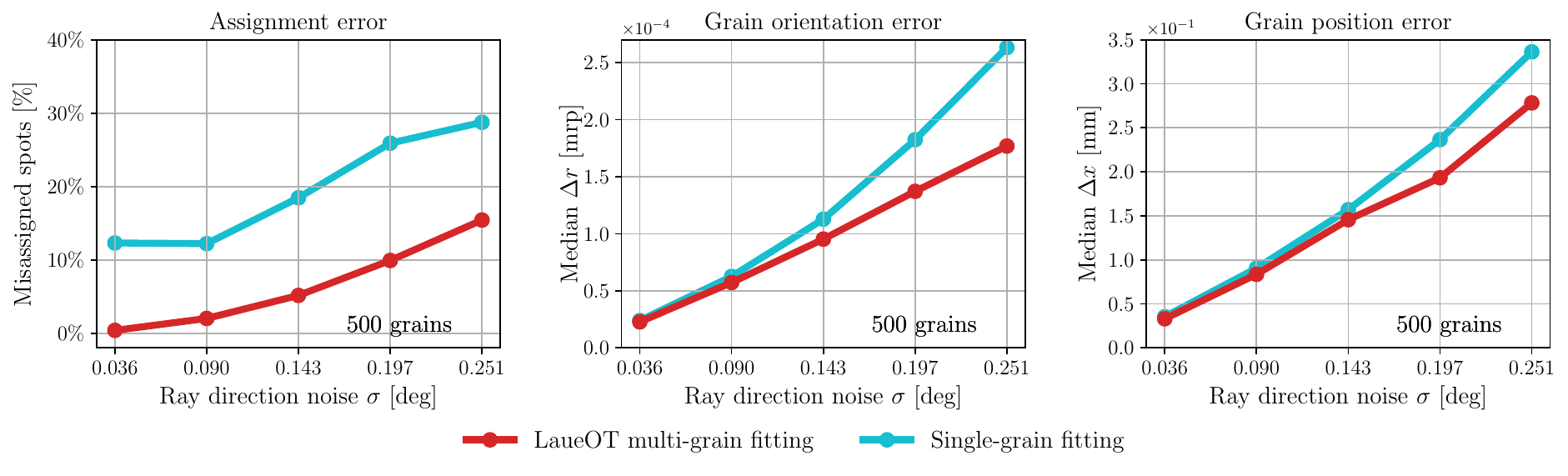}
\caption{Quality metrics, as in Figure~\ref{fig:quality_metrics}, for the multi-grain fitting solution of the full \ours\ solver, compared to the single-grain fitting initializations.
The comparison is done for a fixed number of 500 input grains and the fiducial outlier fraction.
The \ours\ solver finds all grains successfully, while the single-grain fitting results in $\sim$940 correctly indexed grains, with slight dependence on the noise level.}
\label{fig:ot_vs_prototypes}
\end{figure*}

We evaluate our proposed \ours\ algorithm according to the following performance criteria
\begin{enumerate}
  \item accuracy of the solution as a function of number of grains in the sample and the noise level,
  \item sensitivity of the solutions to the fraction of outliers for different optimal transport formulations, and its dependence on OT outlier control parameters,
  \item number of successfully indexed grains as a function of the number of initial single-grain optimizations for two types of lattice,
  \item improvement of the multi-grain solver solution over the single-grain fitting solutions,
  \item comparison of time-to-solution between \ours\ and the \textsc{Laue3DNDT} method from \cite{Raventos2019laue3dndt},
  \item comparison of the indexing results on spots obtained from experimental images.
\end{enumerate}
For simulated data, we use two types of lattices, FeNiMn and CoNiGa, which are metal alloys (see \cite{Boyer1985metals} for introduction to metal alloys).
We test the accuracy of the \ours\ solutions using simulated data in the following way.
We consider a neutron scattering experiment scenario with the wavelength of the beam $\lambda^{} \in [0.6, 6] \ \angstrom$ and two detectors placed at \mbox{$\gamma^{+} = (-160, 0, 0)$} and \mbox{$\gamma^{-}=(160, 0, 0)$} mm.
The fiducial experiment configuration for both samples is 12 projections ($\omega = 0^\circ$\,--\,$360^\circ, \Delta\omega = 30^\circ$).
The noise level for the fiducial configuration corresponds to the scatter of $\sigma{=}0.4$ mm on the center of the detector screen, which is equivalent to angular spread of $\sigma = 0.143$ deg for rays.
The chosen noise level is already conservative, as the size of the point spread function on the scintillator detectors, such as the ones used in neutron scattering studies, is $\sim$0.2 mm, and its pixel size is 0.103 mm.
We add outliers to the simulated detected spot set.
The position of the outliers is drawn from an uniform distribution on the image plane.
The number of outliers in the fiducial configuration is set to 10\% of the Laue spot count.
In the fiducial configuration, the simulated samples contain up to 1000 grains of orientations chosen on a Sobol sequence with additional random scatter. 
 
For synthetic data, we simulated two datasets using a $Fm\bar3m$ space group and a $Pm\bar3m$ space group with lattice parameters $a{=}b{=}c{=}3.592$\,\AA\ and $a{=}b{=}c{=}2.872$\,\AA, respectively. 
Hereafter, the simulated samples will be referred to as SynthSampleA and SynthSampleB, respectively. 
The choice of the above space groups and corresponding lattice parameters was made because they match recent experimental studies \cite{samothrakitis2020multiscale, samothrakitis2022microstructural}. 
For the experimental data, we used a ruby single crystal, typically used in crystallography for calibration measurements, a CoNiGa oligocrystalline sample, and a FeGa polycrystalline sample. Sample properties, including space groups and lattice parameters, are listed in Appendix \ref{app:experimental_sample}. 

Use 32 nearest neighbors in the sparse formulation.
The initial temperature $\epsilon^{\rm{init}}$ is related to the number of nearest neighbors considered in the sparse formulation:
it is chosen so that the farthest neighbor has a non-negligible probability at start.
The temperature is lowered by $\epsilon^{\rm{cool}}{=}0.97$ at every iteration.
For the likelihood of outliers, we use the log likelihood $L^{\rm{out}}$ set to the likelihood corresponding to a angular difference of $3\sigma$.
To discard badly fitting grains, we require the number of rays matched to the outlier to be less than $f^{\rm{out}} \cdot |\mathcal{S}_n|$, with typical value $f^{\rm{out}}{=}0.1$.
For the multi-grain fitting, we set the maximum number of unmatched model rays $T_b = \lceil 0.4 \cdot |\mathcal{S}_n| \rceil$ of total number of rays.

\subsection{Quality metrics}

To assess the quality of the \ours\ solutions, we compare: 
(i) the fraction of correctly assigned spots across all grains, 
(ii) the median difference between true and estimated grain orientation,
(iii) the median difference between true and estimated grain position.
The results of these comparisons are shown in Figure~\ref{fig:quality_metrics}, as a function of number of grains and spot position noise level.

For samples with up to 500 grains, the number of recovered grains was exactly correct. 
After that, there was at most 1 grain missing.
Starting with samples with $700$ grains and for higher noise levels, the method begins to fail.
This gives an idea for the limitations of \ours.
The left panel shows the fraction of wrongly assigned spots as a function of the number of grains and the noise level.
\ours\ mismatches only up to 15\% of spots even for the challenging scenario of 500 grains and high noise level $\sigma{=}0.251^{\circ
}$ ($0.7$ mm on the image plane).
For the low noise level of $\sigma{=}0.036$ deg, \ours\ assigns almost all spots correctly.
As expected, this fraction decreases with increased noise level.
Middle and left panels show the median error of the recovered grain orientation and position, respectively.
Predictably, these errors increase with the noise level.
They stay almost constant with increased number of grains, which is expected in case of good quality of assignments.

Figure~\ref{fig:ot_vs_prototypes} shows the benefit of obtaining solutions by solving the full inverse problem in Equation~\ref{eqn:elbo} with OT, compared to using the prototypes from single-grain fitting only.
It shows the quality metrics, as in Figure~\ref{fig:quality_metrics}, for the SynthSampleA with 500 grains, at varying noise levels and fixed outlier fraction of 10\%.
To assess the performance for the single-grain fitting, a model spot is hard-assigned to the nearest detected spot.
Rays were marked as unmatched when the probability of the match was lower than the vMF likelihood $L^{\rm{out}}$ corresponding to the angle of $3\sigma$.
The OT formulation decreases the fraction of wrongly assigned spots by a factor of 2.
The grain orientation and position errors also improve with the multi-grain fitting, with errors decreasing by up to $25\%$ for high noise levels.

\subsection{Robustness to outliers}
\label{sec:outliers}

\begin{figure*}
\centering
\includegraphics[width=1\linewidth]{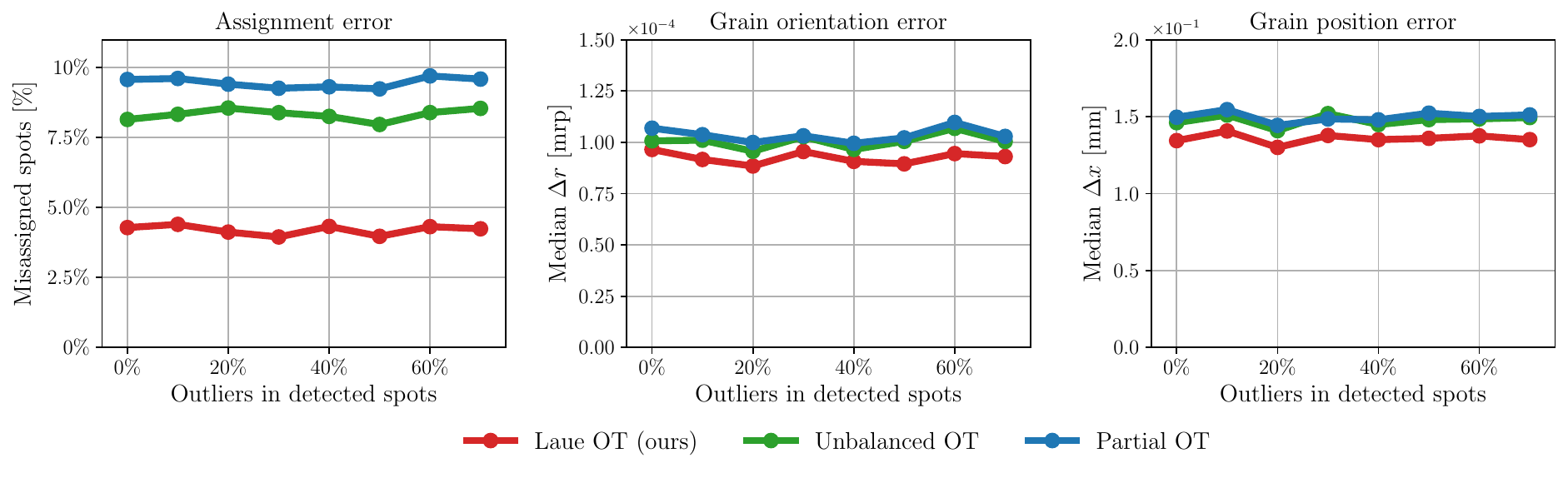}
\caption{Performance of the \ours\ solver for number of outliers in the data, for three outlier treatment methods used.
The hyperparameters for each method were found by performing a grid search over multiple combinations and choosing the one that minimizes the average assignment error over outlier fractions.
}
\label{fig:outliers}
\end{figure*}

We test the sensitivity of the \ours\ method to outliers in the detected spots.
We compare the performance of our proposed \ours\ formulation using probabilistic outlier model (see Section~\ref{sec:problem_with_outliers}), as well as two other OT formulations: partial OT and unbalanced OT (see Appendix~\ref{app:other_ot_methods}).
In this ablation study we explore how this choice affects the indexing results.
In particular, we test the sensitivity of the results to the choices of outlier control parameters: likelihood associated with outliers, maximum number of unmatched rays for OT with outlier modelling, $\kappa$ and $\lambda$ entropic regularization parameters for the unbalanced OT, and fraction of transported mass $m_{\rm{OT}}$ for the partial OT.

We choose simulated diffraction images of the \mbox{SynthSampleA} with noise level \mbox{$\sigma{=}0.143^{\circ}$} as the reference.
We vary the fraction of outliers up to 70\%.
For \ours, we use combinations of outlier cost $L^{\rm{out}}$ and maximum number of unmatched rays $T_b$.
The outlier likelihood corresponds to vMF probability for angular difference of \mbox{2--5$\sigma$}, while the fraction of unmatched rays between \mbox{1--40\%}.
For \textsc{UnbalancedOT}, we set $\kappa,\lambda{\in}[0.001, 1]$.
For \textsc{PartialOT}, the fraction of transported mass is varied between \mbox{$m_{\mathrm{OT}}\in [0.4, 0.95]$}.
For each OT method, we choose the parameter configuration that yields the smallest average assignment error over all outlier fractions.

Figure~\ref{fig:outliers} shows the assignment error (left panel), grain orientation error (middle panel), and position error (right panel), for the three OT outlier treatment methods.
The results indicate that the methods are not affected by outliers.
The best overall results are obtained with the custom outlier modelling in \ours.
The unbalanced and partial OT give slightly worse, but comparable results.
Given that the multi-grain fitting is very fast, the user can perform a coarse grid search to find reasonable values for the hyper-parameters.

\subsection{Number of initializations and runtime}

\begin{figure}
\centering
\includegraphics[width=1\linewidth]{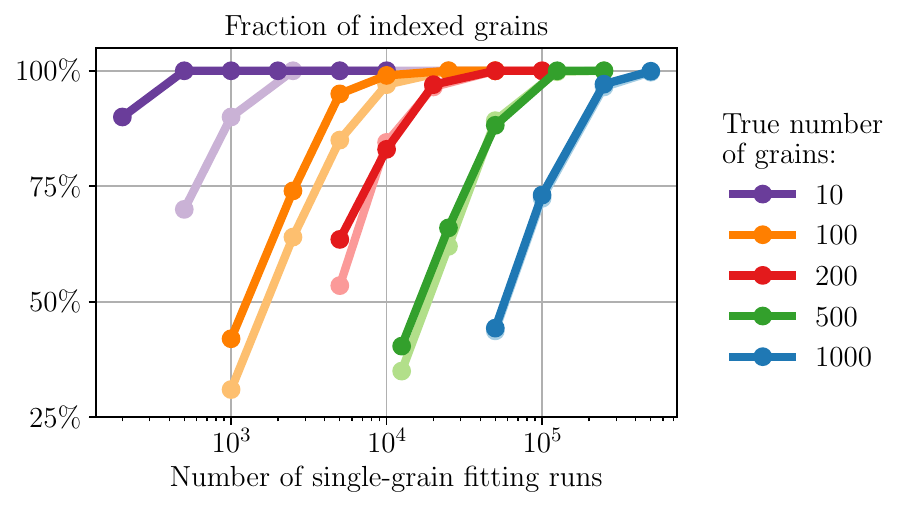}
\caption{Fraction of correctly indexed grains as the number of single-grain fitting initializations considered in the prototype selection, for samples SynthSample1 (darker-colored lines) and SynthSample2 (lighter-colored lines).
The results are shown for varying number of true grains in the simulation, with spot noise level kept constant.}
\label{fig:timings}
\end{figure}

Figure~\ref{fig:timings} shows the number of correctly indexed grains as a function of the number of single-grain fitting solutions considered in the prototype selection.
Different number of true grains in the sample is considered, using fiducial noise level \mbox{$\sigma{=}0.143$} deg.
In this test we considered both SynthSampleA and SynthSampleB.
As expected, more initializations are needed for high number of grains. 

We compare the performance between \ours\ and the \textsc{Laue3DNDT} method from \cite{Raventos2019laue3dndt}, which is currently used for wide-beam Laue tomography problems.
Table~\ref{tab:comparison_with_findpeaks} shows the comparison between the methods limited to results that can be obtained within 24 hours.
This comparison is performed for SynthSampleA using the fiducial noise and outliers, with 32 rotation angles.
In this test, \ours\ finds the exact number of grains within 1h on a single large memory GPU, while \textsc{Laue3DNDT} outputs 356/500 grains within 24 hours.
Further acceleration for \ours\ can be achieved by running on multiple GPUs, as the single-grain fitting is trivially parallelizable.

\subsection{Application to experimental data}
\label{sec:expermiental}

\begin{figure*}
\centering
\includegraphics[width=1\linewidth]{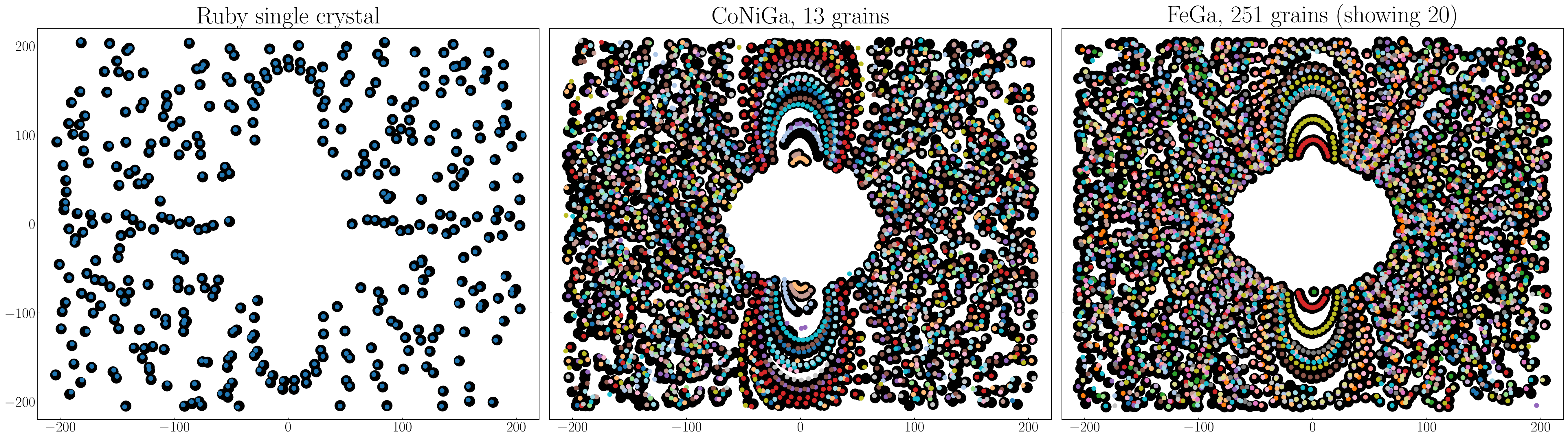}
\caption{Results of \ours\ indexing for three samples: a ruby single crystal (left panel), an oligocrystalline CoNiGa alloy (middle panel), and a polycrystalline FeGa alloy (right panel).
The panels show spots detected at all projection angles stacked together, for the forward-scatter detector.
For FeGa, we only show the matches for the first 20 grains.
The black spots are detected from experimental images, while the colored spots show the spots for the fitted model grains.
Only matched spots are shown; outliers are omitted for clarity.
The statistics of the matched spots are described in Table~\ref{tab:experimental}.
}
\label{fig:experimental}
\end{figure*}

\begin{table*}
\renewcommand{\arraystretch}{1.3}
\label{tab:experimental}
\caption{Crystal grain indexing results for three samples. The numbers in brackets correspond to spot counts in the (backscatter, forward-scatter) detectors.} 
\centering
\begin{tabular}{|c|c|c|c|c|c|}
\hline
Sample       &  Number of projections & Wavelength $\lambda\ [\angstrom]$ & Number of spots      & Number of grains &  Number of matched spots \\
\hline
Ruby         & 46        & [1.0-5.0] & 3512 (3093, 419)        &  1               & 2852 (2468, 384)                     \\
CoNiGa       & 87        & [1.0-5.0] & 16822 (11246, 5576)     &  13              & 11883 (9039, 2844)                   \\
FeGa         & 86        & [1.1-5.0] & 269181 (197047, 72134)  &  251             & 214471 (169240, 45231)               \\
\hline
\end{tabular}
\end{table*}

\begin{table*}
\renewcommand{\arraystretch}{1.3}
\label{tab:comparison_with_findpeaks}
\caption{Time to solution for the \ours\ algorithm compared to the solution from \textsc{Laue3DNDT} \cite{Raventos2019laue3dndt}.} 
\centering
\begin{tabular}{|c||c|c|c|}
\hline
           & sample                    &  number of grains found & time-to-solution\\
\hline
\textsc{Laue3DNDT} & $N^{\mathrm{grains}}=200$ & 196 & 24h \\
\ours              & $N^{\mathrm{grains}}=200$ & 200 & 8 min on a single A100 \\
\textsc{Laue3DNDT} & $N^{\mathrm{grains}}=500$ & 356 & 24h \\
\ours              & $N^{\mathrm{grains}}=500$ & 500 & 30 min on a single A100, 5m on 8$\times$A100 \\
\hline
\end{tabular}
\end{table*}

We demonstrate the performance of the \ours\ solver on spot data taken during neutron diffraction experiments.
We show results on three samples: 
(1) a ruby single-grain calibration sample, 
(2) a CoNiGa alloy oligocrystal, and 
(3) a FeGa alloy polycrystal.
The details of the experiments are described in Appendix~\ref{app:experimental}.
Table~\ref{tab:experimental} describes the samples in terms of the number of projection angles used in the experiment, the number of spots detected from the images, and the wavelength range used in the \ours\ algorithm. 
The number of grains found by \ours\ is 1, 13, and 251 for the ruby, CoNiGa, and FeGa, respectively. 
The runtime for this analysis was few minutes for the ruby, 5 min for CoNiGa and $30$ min for FeGa, on a single A100 GPU.
The number of matched spots is 70-80\%.
The orientation and position of the ruby crystal found by \ours\ agrees well with analysis by \textsc{Laue3DNDT}.
The ruby crystal measurements was used to calibrate the experimental setup in terms of distances between the detectors and the sample, as well as tilts of the detector plane \cite{samothrakitis2023calibr}.

Figure~\ref{fig:experimental} shows the results of indexing of the ruby, CoNiGa, and FeGa samples in the left, middle, and right panels, respectively.
The panels show stacked spots for all projection angles in the forward-scatter detector.
As in Figure~\ref{fig:steps}, the black points show the spots detected in the experiment, and the colored points show the model fitted by \ours, with different colors corresponding to model grains.
For clarity, only the matched spots are shown, the outliers are omitted.
For the FeGa sample, we only show spots for the first 20 grains.
The empty area in the middle is due to the fact that rays close to direction of the beam can not satisfy the Bragg condition (Equation~\ref{eqn:bragg}).
The agreement is very good for the ruby, and good for the CoNiGa and FeGa samples, up to the precision level of calibrations.

\section{Conclusions}
\label{sec:conclusions}

We presented a novel approach to indexing grains in multi-crystal samples from wide-beam diffraction experiments.
Our method, called \ours, poses a principled optimization problem formulation for the grain indexing.
The problem is solved using a computationally efficient coordinate descent method, with assignment of grains--to--spots performed in the framework of optimal transport. 
A crucial part of this model is the treatment of outliers in the experimental data and of redundant model spots.
To treat the outliers we propose a model dedicated to the Laue indexing problem, based on a full probabilistic description of outliers.
Our new probabilistic model clearly outperforms the two existing formulations in the literature, partial and unbalanced OT.

We show the performance of \ours\ using simulations for variable number of grains in the sample, as well as different noise levels and outlier fractions.
For samples with $<500$, the method recovers all the input grains.
For larger samples, the method occasionally misses a single grain.
We found the limit of the stable performance of the algorithm is reached for samples with large number of grains at high noise levels.
The accuracy of the fitted grain positions and orientations is generally very good, but decreases slightly with increased noise level.

The \ours\ methods opens the possibility of analyzing samples with large number of grains with short time-to-solution.
For samples with $\sim$500 grains, the time-to-solution \ours\ on a Nvidia DGX node ($8 \times$ A100 GPUs) is of order few minutes, compared to $>$24 hours taken by the original indexing method of \textsc{Laue3DNDT}.
This enables almost real-time sample analysis during the experiment, as well as simulation-based setup optimization before starting the measurements.

\appendices

\section{Wahba's problem}
\label{app:wahba}

For two sets of points and their correspondences, the rotation matrix can be found by solving the Orthogonal Procrustes problem.
This problem is solved through the singular vector decomposition.
\emph{Wahba's} problem \cite{Wahba1965satellite} generalizes that to the analysis of points with weights and addresses the sign ambiguity of the SVD output.
The rotation matrix $R$ for a fixed set of $K$ pair of 
points $a_k$ and $b_k$ and the assignment weight $w_k$ using the subroutine \textbf{Wahba} (Algorithm~\ref{alg:wahba}).
Similarly to the Orthogonal Procrustes problem, it uses the singular value decomposition, but additionally uses a scaling by the product of determinants to remove the sign ambiguity in the SVD.

\section{M-steps for grain position and rotation}
\label{app:mstep}

In this section we derive the M-step updates for grain position $x$ and rotation $R$, as introduced in Section~\ref{sec:fitting_rot_pos}.
For both parameters, we find the updates corresponding to the maximum of the likelihood in Equation~\ref{eqn:elbo}.
The M-step requires solving the optimization problem for position $x_n$ and $R_n$ of every grain $n\in N$ separately 

\begin{align}
    \forall n \in N : \ \displaystyle{\min_{x_n, R_n}} \quad \mathcal{L}(R_n, x_n) = \frac{1}{\sigma^2} \sum_{ij} Q_{ij} \ \hat r_{ij}\T r_{j}.
\end{align}
We perform coordinate descent alternating between updates for $R$ and $x$.

\subsection{Grain position}

We solve this problem for $x$ using the Newton's method. The gradient $\nabla \mathcal{L}(x)$  and Hessian $\nabla^2 \mathcal{L}(x)$ are  
{\scriptsize
\begin{flalign}
    \nabla \mathcal{L}(x) &=  \frac{1}{\sigma^2} \sum_{ij} \frac{Q_{ij}}{\tau_{ij}} r_{j}\T \left( \mathbbm{I} - \hat r_{ij} \hat r^{\top}_{ij} \right), \label{eqn:xgrad} &&\\
    \nabla^2 \mathcal{L}(x) &=  \frac{1}{\sigma^2} \sum_{ij} \frac{Q_{ij}}{\tau_{ij}^2} \left(    3 \cdot r_{j}\T \hat r_{ij} \cdot  \hat r_{ij} \hat r_{ij}\T - 
      r_{j} \hat r_{ij}\T - \hat r_{ij} r_{j} \T - r_{j}\T \hat r_{ij} \cdot \mathbb{I}     \right), \label{eqn:xhess} &&
\end{flalign}
}%
where $\tau_{ij}{:=}\| s_i - \Gamma_{t_j} x_{n_j} \| $ is the length of the ray $\hat r_{ij}$ from the grain center to the spot position $s_i$ on the screen.
The Netwon's method finds the step $\Delta x$ by solving \mbox{$\Delta x \leftarrow [\nabla^2 \mathcal{L}(x)]^{-1} \nabla \mathcal{L}(x)$}.
Subroutine \textbf{LauePosNewton} in Algorithm~\ref{alg:posnewton} describes the details of this process.
The iterations stop when a convergence threshold $\zeta^{\mathrm{tol}}$ on position is reached.

\subsection{Grain rotation}

As described in Section~\ref{sec:fitting_rot_pos}, finding $R$ requires solving a quadratic programming problem with orthogonality constraints.
In this section we derive the \textbf{LaueRotMM} algorithm, which finds the maximum likelihood solution.
The optimization problem in Equation~\ref{eqn:elbo} is equivalent to
\begin{align}
\begin{aligned}
\label{eqn:rot_mirror1}
\min_{R} - \sum_{j \in J_n} \Big \langle \sum_{i \in I} Q_{ij} \hat{r}_{ij}, \ \Gamma_{t_j} R w_{m_j} w_{m_j}\T R\T \Gamma_{t_j}\T e \Big \rangle,\\
\end{aligned}
\end{align}
such that $R^\top R=\mathbbm{I}$ and where we dropped the index of the grain from the rotation matrix $R.$
By introducing the matrix 
\begin{align}
\begin{aligned}
M^0_m:=- \sum_{j \in J_n \cap J_m } \Gamma_{t_j}^\top \Big [ \sum_{i\in I} Q_{ij} \, &( \hat{r}_{ij}e^\top + e \hat{r}_{ij}^\top ) \Big ] \Gamma_{t_j},
\end{aligned}
\label{eqn:M0}
\end{align}
this problem can be expressed as 
\begin{align}
\label{eqn:qcoc}
& \min_{R: R^\top R= \mathbbm{I}} -\sum_{m \in M}  w_m^\top \, R^\top M^0_m \, R\,  w_m.
\end{align}
The matrices $M_m^0$ are generally not positive semi-definite, which makes the problem not concave.
By using a positive semi-definite matrix 
\begin{align}
\label{eqn:Mm}
M_m:=M_m^0+|\iota_m| \mathbbm{I}, 
\end{align}
where $|\iota_m|$ is the smallest eigenvalue or $0$ in case $M_m^0$ has no negative eigenvalue, we can write an equivalent optimization problem
\begin{align}
\label{eqn:qcoc_concave}
& \min_{R: R^\top R=\mathbbm{I}} -\sum_m  w_m^\top \, R^\top M_m \, R\,  w_m.
\end{align}
This problem is the Quadratic Program with Orthogonality Constraints (QP-OC). 
There exist multiple solvers for this problem in literature \cite{ManSo2009qpoc,Liu2016quadratic}.
Here we design a dedicated Majorization Minimization method using an upper bound on the function in Equation~\ref{eqn:qcoc}.

The associated quadratic is an upper bound of its first order Taylor expansion around a rotation matrix $R^k$
and so 
\begin{equation}
\begin{split}
-&\sum_m  w_m^\top \, R^\top M_m \, R\,  w_m  \\
  &\leq   \sum_m  w_m^\top \, R^{k\top} M_m \, R^k\,  w_m- \sum_{m} 2 \, w_m^\top \, R^{k\top} M_m \, R\,  w_m .
\end{split}
\end{equation}
The minimization of the upper bound is a Wahba's problem (see Appendix~\ref{app:wahba}) with unity weights:
\begin{align}
& \min_{R: R^\top R=\mathbbm{I}} - \sum_{m} \, w_m^\top \, R^{k\top} M_m \, R\,  w_m\\
\Leftrightarrow & \min_{R: R^\top R=\mathbbm{I}} \sum_{m} \|R\,  w_m- M_m R^k w_m\|^2.\label{eqn:majmin_step}
\end{align}
The Minimization-Majorization scheme starts with an initial guess $R^0$ and performs a number of iterative steps according to Equation~\ref{eqn:majmin_step}.
The routine \textbf{LaueRotMM} is shown in Algorithm~\ref{alg:rotmajmin}.
The iterative steps are performed until a given convergence thresholds on rotation $\zeta^{\mathrm{tol}}$ is reached.

\section{Optimal transport variants}
\label{app:other_ot_methods}

As described in Section~\ref{sec:problem_with_outliers}, our main proposed method for finding the transport plan between the model and detected spots is using a custom probabilistic model for outlier treatment. 
We compare the performance of this model to other OT variants that allow for outlier treatment: unbalanced and partial optimal transport. 
For all our Sinkhorn methods, the convergence is controlled by a maximum difference to previous iteration $\zeta^{\mathrm{tol}}$.

\subsection{Partial optimal transport}
\label{sec:partial_ot}

Partial optimal transport \cite{Figalli2010partial} aims to move only a pre-defined fraction $m_{\mathrm{OT}}$ of the mass ---here the number of spots--- that has to be transported in the OT formulation.
The optimization problem has a single control parameter, which is specified by the user.
The objective function for finding the transportation plan $Q$ is
\begin{equation}\tag{A}
\begin{aligned}\label{eqn:elbo3}
\displaystyle{\max_{Q}} \quad &\left\langle L, Q \right\rangle + \epsilon \mathbf{H}(Q)   \\[0.5em]
s.t.                          \quad & Q \geq 0, \quad Q \in \mathcal{Q}\\
                              \quad & Q\mathbbm{1} \leq a, \quad Q\T\mathbbm{1} \leq b, \quad \mathbbm{1}\T Q\T \mathbbm{1} = m_{\mathrm{OT}} ,
\end{aligned}
\end{equation}
where $a$ and $b$ are vectors corresponding to spot weights (typically equal and set to one),
and \mbox{$m_{\mathrm{OT}} \leq \min \{ \|a\|_1, \|b\|_1 \}$}, and $\epsilon$ is the temperature parameter (see Section~\ref{sec:annealing}), which should be set to $\epsilon{=}1$ to match Equation~\ref{eqn:likelihood}.
In partial OT, the spots to be matched are selected automatically based on their cost. 
A motivation to consider Partial OT in the presence of outliers and missing spots is that one can hope that only inliers will be matched with their corresponding models spots. 

\subsection{Unbalanced optimal transport}
In this formulation, the constraint that the mass distribution in $Q$ should have marginals that are equal to $a$ and $b$ is relaxed, and they are only encouraged to be close to them by the means of regularization, typically a generalized KL-divergence regularizer due to availability of closed-form solution \cite{Chizat2018scaling}.
The optimization problem for transportation plan $Q$ in Problem (B) is 
\begin{equation}\tag{B}
\begin{aligned}\label{eqn:opt_unbalanced}
\displaystyle{\max_{Q}} \quad &\left\langle L, Q \right\rangle + \epsilon \mathbf{H}(Q) - \kappa \mathbf{D}(Q\mathbbm{1}, a) -  \lambda \mathbf{D}(Q\T\mathbbm{1}, b)   \\
s.t.                          \quad &  Q \geq 0, \quad Q \in \mathcal{Q}
\end{aligned}
\end{equation}
where the proximity term $\mathbf{D}(x,y)$ is the generalized Kullback-Leibler divergence between $x$ and $y$, 
\mbox{  $\mathbf{D}(x, y) = x\T \log\left( x \oslash y \right)  + (y-x)\T\mathbbm{1}$ },
where $\oslash$ is the element-wise division, and $\epsilon$ is the entropic regularization parameter.
The strength of these marginal constraint regularization terms controlled by $\kappa$ and $\lambda$.

In practice, due to the entropic term, a consequence of relaxing the equality to the marginal, is that, as in partial OT, the mass transported is smaller than the total mass, which motivates the possible use of this method in the presence of outliers and missing spots.
Here, $\kappa$ and $\lambda$ influence the outliers in source and target distributions, which will be ignored if the cost of transporting them is too high.
In the limiting case of of $\kappa \rightarrow \infty, \lambda \rightarrow \infty$, the solution converges to the balanced optimal transport where no outliers are allowed.

\section{Single-grain fitting}
\label{app:single_grain_fitting}

The single-grain fitting is performed using the simplified version of \ours.
The key difference is that we remove the double-stochastic constraint, which does not make sense for a single grain.
The reason for this is that the model spots from a single grain are sufficiently far from each other, that it is extremely unlikely that they would be matched to a common detected spot.
Furthermore, removing this constraint enables us to perform assignment updates in a single step, without having to solving Sinkhorn.
This speeds up the process significantly.
We employ the deterministic annealing scheme here too to obtain better local minima.
We maximize the function
\begin{equation}
\begin{aligned}\label{eqn:elbo_single}
\displaystyle{\max_{Q, R, x}} \quad &\left\langle L, Q \right\rangle + \epsilon \mathbf{H}(Q)  \\
s.t.                          \quad &  Q \geq 0, \quad Q \in \mathcal{Q}, \quad Q\mathbbm{1} = a, \quad  R\T R = \mathbb{I}, \\
\end{aligned}
\end{equation}
where $L$ is the likelihood matrix in Equation~\ref{eqn:update_L}.
With the double-stochastic constraint removed, the assignment matrix is computed in a single update.
Algorithm~\ref{alg:single} for single-grain fitting is described in detail in Appendix~\ref{app:single_grain_fitting}.
We perform the optimization for model spots generated from a single grain model.
The annealing schedule is following the same rules as the multi-grain fitting.
Here we set the number of iterations to a fixed number $k^{\mathrm{max}}$, which allows us to use perform multiple single-grain fitting runs in a batch mode. 
This leads to a significant computational performance boost.
For grain position and rotation updates, we perform 4 steps in the Newton and MM iterations.
This is done to speed up the computation, as the single-grain fitting prioritizes speed over precision.

\section{Selection of prototypes with OT}
\label{app:proto}

Let's consider a set of candidate grains
\mbox{$\{ \mathcal{G}^{\rm{c}}_n \}_{n \in N^{\rm{c}}}$}, indexed by $N^{\rm{c}}$, which are solutions obtained by single-grain fitting, that have the number of outliers $ < f^{\rm{out}} |\mathcal{S}_n|$.
Each candidate will create model spots indexed by $J^{\rm{c}}_n$.
The goal is to find a subset of initial grains $N^{\mathrm{p}} \subseteq N^{\rm{c}}$, called the \emph{prototypes}, that will constitute the final sample.
We start with calculating the grain-to-spot assignment log-likelihood matrix $\bar L \in \mathbbm{R}^{|I| \times |N^{\rm{c}}|} $.
The cost of assignment of spot $s_i$ to grain $\mathcal{G}_n^{\rm{c}}$ is calculated using the marginal probability of assignment for all model spots in the grain
\begin{align}
  \bar L_{in} & \leftarrow  \log \sum_{j \in J^{\rm{c}}_n} p(s_{i}, Z_{ij}=1).
\end{align}
We then shift the cost matrix to positive values 
\begin{align}
  \bar L_{in}  &\leftarrow  \| \bar L \|_{\infty} + \bar L_{in}.
\end{align}
The prototype selection depends on finding a weight vector \mbox{$w \in \Delta_{N^{\rm{c}}}$} corresponding to candidate grains, where simplex $\Delta_{N^{\rm{c}}} := \{z \in \mathbb{R}_{+}^{N} \ | \ z\T \mathbbm{1} = 1  \}$.
We define a set function

\begin{equation}
\begin{aligned}\label{eqn:prototype_function}
f(N^{\rm{p}}) &:= \max_{w: \ supp(w) \subseteq  N^{\rm{p}}}  \ \max_{P} \langle \bar L,  P \rangle \\
s.t.                          \quad &  P \geq 0, \quad P\mathbbm{1} = a, \quad P\T \mathbbm{1} = w,
\end{aligned}
\end{equation}
where $P$ is a OT assignment matrix and $f(\emptyset)=0$.
The objective function finds $N^{*}$ that maximizes $f$ subject to cardinality constraint on \mbox{$|N^{*}| < k$}, enforcing that the size of the set is smaller than $k$.
The \spot\ loss is
\begin{align}\label{eqn:prototype_loss}
  N^{*} = \argmax_{N^{\rm{p}} \subseteq N^{c}, |N^{\rm{p}}| \leq k} f(N^{\rm{p}}). 
\end{align}
Finding the optimum subset $N^{*}$ is NP complete, but can be solved approximately using greedy incremental methods, which leverage its submodular property.
The algorithm proposed in \cite{Gurumoorthy2021spot} has proven approximation guarantees \cite{Nemhauser1978submodular}.

We employ the \textsc{SpotGreedy} algorithm \cite{Gurumoorthy2021spot} to find the approximate solution for Equation~\ref{eqn:prototype_loss}.  
This algorithm selects prototypes in a greedy fashion, starting with the one that contributes the most to the objective function.
The relative contributions of consecutive prototypes to the objective reflect the degree of uniqueness of the prototype.
When the increment in the cost functions for a new candidate prototype is small, that indicates that it is similar to a prototype that already has been accepted.
Due the greedy nature of the algorithm, for cardinality $k+1$ it will output the same elements in the same order as for $k$, extended by one new element.

Practically, the \textsc{SpotGreedy} algorithm starts with constant 
outputs a sequence of indices and their corresponding objective function values $(n_k, f_k)_{k \in N^{\rm{*}}}$, indexed by $N^{\rm{*}}$.
A subsequence with elements up to $k$ corresponds to the prototypes $N^{\rm{p}} \subseteq N^{c}$, for cardinality constraint $|N^{\rm{p}}| \leq k$.
To choose the final set of prototypes $N^{\rm{p}}$ with size $k^{*}$, we employ a criterion based on 2nd discrete difference of $f_{k}$
\begin{equation}
\textstyle
\label{eqn:spot_ngrains}
  k^{*} \leftarrow \argmax_k  f_{k+1}+f_{k-1}-2f_{k}
\end{equation}
for $k>1$.
We use a fractional criterion instead of the commonly used cut-off on the objective increment $\epsilon$ (see Algorithm~1 in~\cite{Gurumoorthy2021spot}), as it requires no knowledge of the numerical values of the objective from the user.
We call this method \mbox{\textbf{LaueSpotGreedy}} (Algorithm~\ref{alg:prototypes}) and describe it in Appendix~\ref{app:proto}.
The prototype grain set is passed on to the next step of the algorithm, which solves problem in Equation~\ref{eqn:elbo}.
This procedure typically finds the correct number of grains, except when the number of candidates obtained from single-grain fitting is not sufficiently large.
The number of single-grain fitting candidates required is described in Section~\ref{sec:experiments}.

\section{Experimental details}
\label{app:experimental}

\subsection{Samples}
\label{app:experimental_sample}

\textbf{Ruby:} As a first sample, a Ø 6\,mm spherical single-crystal ruby, typically used for detector calibrations, was measured. For the indexing of the ruby crystal, a $R\bar{3}c$ space group with lattice parameters of $a{=}b{=}4.7606$~\AA, $c{=}12.994$~\AA\ was used. \\ 

\noindent \textbf{CoNiGa:} A cylindrical Ø 4\,mm\,$\times$\,4\,mm oligocrystalline CoNiGa sample was selected as an intermediate test case. 
The sample was prepared via hot-extrusion followed by post-extrusion heat treatment, resulting in millimeter-sized grains. 
The full details of the sample preparation are given in~\cite{samothrakitis2020multiscale}. 
For the indexing of the CoNiGa sample a cubic $Pm\bar{3}m$ space group with lattice parameters of $a{=}b{=}c{=}2.8657$~\AA\ was used. \\

\noindent\textbf{FeGa:} As a final test case, a $3\times3\times3$\,mm cube Fe$_{0.9}$Ga$_{0.1}$ polycrystalline sample was measured. 
The sample was prepared via arc-melting of 99.95\% purity Fe and Ga buttons in an argon atmosphere. 
To obtain a homogenous distribution of elements, the sample was re-melted 4 times. 
After arc-melting, the sample was annealed in argon at a temperature of 1000\,$^\circ$C with a heating rate of about 20\,$^\circ$C/min and a holding time of 12\,h with subsequent quenching to room temperature. 
The final sample shape was achieved via spark cutting and subsequent polishing with sandpaper, diamond suspension, and colloidal silica. 
For the indexing of the FeGa data set, a $Im\bar{3}m$ space group with lattice parameters of \mbox{$a{=}b {=}c{=}2.917$~\AA\ } were used.

The noise level was set to $\sigma{=}1.07^{\circ}$ and $\sigma{=}0.27^{\circ}$  for CoNiGa and FeGa samples, respectively. 
The CoNiGa sample required higher noise level, as the uncertainty also included systematic errors in the detector setup calibration.
For both CoNiGa and FeGa samples, the number of spots detected in each rotation varied by up to $40\%$. 
This is not expected, as the number of spots per projection should remain constant.
This large variation can be caused by drops in beam flux for that exposure, as well as failures of the spot detection algorithm.
Some rotation angles were removed from the analysis, as they had abnormally small number of detected spots compared to other angles.
To address this, we set the single-grain selection threshold $f^{\rm{out}}<0.4$ for CoNiGa and $f^{\rm{out}}<0.35$ for FeGa.

\subsection{Neutron Laue diffraction tomography}

All neutron Laue diffraction tomographies were collected at the Pulse Overlap Diffractometer (POLDI) \cite{stuhr2006tof} at the SINQ neutron source of the Paul Scherrer Institute (PSI), Switzerland, using a white beam and the FALCON double-detector system \cite{samothrakitis2023calibr}. 
The FALCON forward- and backscattering detectors were placed upstream and downstream from the sample, respectively. 
Measurement parameters such as the sample-to-detector distances and exposure times were chosen for each sample to optimize signal quality and overall spot coverage on the detector.

The ruby tomography was carried out with sample-to-detector distances of 16.5\,cm  and 16.8\,cm for the back- and forward-scattering detector, respectively. 
The sample was rotated a full 360$^\circ$ in steps of 8$^\circ$, resulting in 46 projections in total. 
Each projection was measured with an exposure time of 90\,s. 

The CoNiGa tomography was carried out with sample-to-detector distances of 16.5\,cm  and 12.5\,cm for the back- and forward-scattering detector, respectively. 
The tomography was performed by rotating the sample a full 360$^\circ$ in steps of 4$^\circ$, resulting in a total of 91 projections. 
Each projection was measured with an exposure time of 90\,s. 

The FeGa tomography was carried out with sample-to-detector distances of 16.5\,cm  and 12.5\,cm for the back- and forward-scattering detector, respectively. 
The tomography was performed over a full 360$^\circ$ rotation in steps of 4$^\circ$, resulting in a total of 91 projections. 
Each projection was measured with an exposure time of 100\,s.

\section{Algorithms}
\label{app:algos}

The symbol $\oslash$ stands for element-wise division.
Element-wise minimum of elements between two vectors $x$ and $y$ is denoted as $\mathbf{min}(x, y)$. 
The $\mathbf{diag}(x)$ function creates a diagonal matrix from an input vector $x$.
Vectors $a$ and $b$ correspond to data point weights, which in \ours\ are set to $a\leftarrow \mathbbm{1}, b\leftarrow \mathbbm{1}$.
\textsc{LaueOT} employs balanced optimal transport with explicit outlier modeling, the iterative updates are shown in Algorithm~\ref{alg:sinkhorn_balanced}.
For \textsc{UnbalancedOT}, the iterative updates are shown in Algorithm~\ref{alg:sinkhorn_unbalanced}.
For \textsc{PartialOT}, the iterative updates are shown in Algorithm~\ref{alg:sinkhorn_partial}.
These algorithms are described in Appendix~\ref{app:other_ot_methods}.

\begin{algorithm}[H]
\caption{\textbf{LauePosNewton}: Newton's method for finding grain positions.}
\label{alg:posnewton}
\begin{algorithmic}[1]
\footnotesize
\State \textbf{input:} $(s_i, w_{m_j}, \Gamma_{t_j})_{j \in J_n}$, $e$
\State \textbf{input:} $x_n^0$, $R_n$, $(Q_{ij})_{j \in J_n, i \in I}$,
\State \textbf{input:} $\zeta^{\mathrm{tol}}$, $k^{\mathrm{max}}$
\State \textbf{set:} $k \leftarrow 0$, \ $\zeta \leftarrow \inf$
\While{ $\zeta > \zeta^{\rm{tol}}$  and $k<k^{\mathrm{max}}$}
        \State $g \leftarrow \nabla \mathcal{L}(x^k)$ (Equation \ref{eqn:xgrad})
        \State $H \leftarrow \nabla^2 \mathcal{L}(x^k)$ (Equation \ref{eqn:xhess})
        \State $\Delta x_n^{k} \leftarrow H^{-1} g$
        \State $x_n^{k+1} \leftarrow x_n^{k} - \Delta x_n^k$
        \State $\zeta \leftarrow \|x_n^{k+1} - x_n^{k}\|_2$
        \State $k \leftarrow k+1$
\EndWhile
\State \textbf{output:} $x_n^k$
\State a
\end{algorithmic}
\end{algorithm}

\begin{algorithm}
\caption{\textbf{Wahba}: solution to the Wahba's problem}
\label{alg:wahba}
\begin{algorithmic}[1]
\footnotesize
\State \textbf{input:} $a$, $b$, $w$ 
\State ${B} \leftarrow \sum _{{k=1}}^{{K}}w_{k}{a}_{k}{{b}_{k}}\T $
\State ${{U}}{{S}}{{V}}\T \leftarrow \mathbf{svd}(B)$
\State $M \leftarrow \operatorname{\mathbf{diag}}[1, \ 1, \ \mathbf{det}({{U}})\cdot \mathbf{det}({{V}})]$
\State ${R} \leftarrow {{U}}{{M}}{{V}}^{T}$
\State \textbf{output:} $R$
\end{algorithmic}
\end{algorithm}

\begin{algorithm}
\caption{\textbf{LaueRotMM}: Majorization-Minimization algorithm for finding rotations with reflected points.}
\label{alg:rotmajmin}
\begin{algorithmic}[1]
\footnotesize
\State \textbf{input:} $(s_i, w_{m_j}, \Gamma_{t_j})_{j \in J_n}$, $e$, $R_n^0$, $x_n$, $(Q_{ij})_{j \in J_n, i \in I}$, $\zeta^{\mathrm{tol}}$, $k^{\mathrm{max}}$
\State \textbf{set:}  $k \leftarrow 0$, $\zeta \leftarrow \inf$
\State \textbf{set:} $M_m \leftarrow M_m^{0}+|\iota_m| \mathbbm{I}$ \quad (Equation~\ref{eqn:Mm}) 
\While{ $\zeta > \zeta^{\rm{tol}}$ and $k<k^{\mathrm{max}}$ }
        \State $R_n^{k+1} \leftarrow \textbf{Wahba}(M_m R^k w_m, w_m, \mathbbm{1}) $
        \State $\zeta \leftarrow \|R_n^{k+1} - R_n^{k}\|_F$
        \State $k \leftarrow k+1$
\EndWhile
\State \textbf{output:} $R_n^k$
\end{algorithmic}
\end{algorithm}

\begin{algorithm}
\caption{\textbf{BalancedOT}: Sinkhorn algorithm for balanced optimal transport with entropic regularization.}
\label{alg:sinkhorn_balanced}
\begin{algorithmic}[1]
\footnotesize
\State \textbf{input:} $L$, $\epsilon$, $\zeta^{\rm{tol}}$
\State \textbf{set:} $K \leftarrow \exp(L/\epsilon)$ , $x^{0} \leftarrow \mathbbm{1}$, $y^{0} \leftarrow \mathbbm{1}$
\State \textbf{set:} $\zeta \leftarrow \inf$, $k \leftarrow 0$
\While{ $\zeta > \zeta^{\rm{tol}}$ }
    \State $x^{k+1} \leftarrow a \oslash(Ky^{k})$
    \State $y^{k+1} \leftarrow b \oslash (K\T x^{k+1})$
    \State $\zeta \leftarrow \max( \|x^{k+1} - x^{k}\|,  \|y^{k+1} - y^{k}\|)$
    \State $k \leftarrow k+1$
\EndWhile
\State $Q \leftarrow \textbf{diag}(x^{k}) \cdot K \cdot \textbf{diag}(y^{k})$
\State \textbf{output:} $Q$ 
\end{algorithmic}
\end{algorithm}

\begin{algorithm}
\caption{\textbf{UnbalancedOT}: Sinkhorn algorithm for unbalanced optimal transport with entropic regularization.}
\label{alg:sinkhorn_unbalanced}
\begin{algorithmic}[1]
\footnotesize
\State \textbf{input:} $L$, $\epsilon$, $\kappa$, $\lambda$, $\zeta^{\rm{tol}}$
\State \textbf{set:} $K \leftarrow \exp(L/\epsilon)$, $x^{0} \leftarrow \mathbbm{1}$, $y^{0} \leftarrow \mathbbm{1}$
\While{ $\zeta > \zeta^{\rm{tol}}$ }
    \State $x^{k+1} \leftarrow ( a \oslash(Ky^{k}))^{\frac{\lambda}{\lambda+\epsilon}}$
    \State $y^{k+1} \leftarrow ( b \oslash (K\T x^{k+1}) )^{\frac{\kappa}{\kappa+\epsilon}}$
    \State $\zeta \leftarrow \max( \|x^{k+1} - x^{k}\|,  \|y^{k+1} - y^\mathrm{k}\|)$
    \State $k \leftarrow k+1$
\EndWhile
\State $Q \leftarrow \textbf{diag}(x^{k}) \cdot K \cdot \textbf{diag}(y^{k})$
\State \textbf{output:} $Q$ 
\end{algorithmic}
\end{algorithm}
\begin{algorithm}
\caption{\textbf{PartialOT}: Sinkhorn algorithm for partial optimal transport with entropic regularization.}
\label{alg:sinkhorn_partial}
\begin{algorithmic}[1]
\footnotesize
\State \textbf{input:} $L$, $\epsilon$, $m_{\mathrm{OT}}$, $\zeta^{\rm{tol}}$
\State \textbf{set:} $K \leftarrow \exp(L/\epsilon)$, $Q_1, Q_2, Q_3 \leftarrow \mathbbm{J}_{IJ}$
\State \textbf{set:} $x \leftarrow \mathbbm{1}$, $y \leftarrow \mathbbm{1}$, $K^{0} \leftarrow K$
\While{ $\zeta > \zeta^{\rm{tol}}$ }
    \State $K_1  \leftarrow \mathbf{diag}[\mathbf{min} ( a \oslash K^{k} \mathbbm{1}, \ x)]  [K^{k} \cdot Q_1]$
    \State $Q_1  \leftarrow Q_1 \cdot K^{k} \oslash K_1 $
    \State $K_2  \leftarrow [K_1 \cdot Q_2] \ \mathbf{diag}[ \mathbf{min} ( b \oslash K_1\T \mathbbm{1}, \ y)]$
    \State $Q_2  \leftarrow Q_2 \cdot  K_1 \oslash K_2 $
    \State $K^{k+1}    \leftarrow K_2 \cdot Q_3 \cdot m_{\mathrm{OT}} / (\mathbbm{1}\T K_2 \mathbbm{1}) $
    \State $Q_3  \leftarrow Q_3 \cdot K_2 \oslash {K^{k}}$
    \State $\zeta \leftarrow \|K - K^\mathrm{prev}\|_F$
    \State $k \leftarrow k+1$
\EndWhile
\State $Q\leftarrow K$
\State \textbf{output:} $Q$ 
\end{algorithmic}
\end{algorithm}

\begin{algorithm}
\caption{\textbf{LaueSpotGreedy}: Selection of grain prototypes using optimal transport in the Laue indexing problem.}
\label{alg:prototypes}
\begin{algorithmic}[1]
\footnotesize
\State \textbf{input data:} $N^{\mathrm{c}}$, $\bar L_{in}$
\State \textbf{initialize:} $N^{0} = \emptyset$
\For{ $ 1 \leq k \leq |N^{\mathrm{c}}|$ }
    \State Define vector $f_{n} \leftarrow f(N^{k-1} \cup N^{\rm{c}}_{n}) - f(N^{k-1})$
    \State Find largest increment $n_k \leftarrow \max_n f_{n}$ 
    \State Create a grain set for $k$: $N^{k} \leftarrow  N^{k-1} \cup N_{n_{k}}^{\rm{c}}$
    \State Store the increment function $f_k \leftarrow f_{n_k}$
\EndFor
\State Get final number of grain prototypes $k^{*}$ (Equation~\ref{eqn:spot_ngrains})  
\State Get the final prototype grain set $N^{*} \leftarrow N^{k^{*}}$
\State \textbf{output}: $N^{*}$ 
\end{algorithmic}
\end{algorithm}

\begin{algorithm}
\caption{\textbf{SingleLaueOT}: Single-grain fitting for Laue indexing.}
\label{alg:single}
\begin{algorithmic}[1]
\footnotesize
\State \textbf{input data:} $(s_i)_{i \in I}$, $\sigma$  
\State \textbf{input number of starting grains:} $N^{\mathrm{s}}$ 
\State \textbf{input outlier parameters:} $L^{\rm{out}}$
\State \textbf{input control parameters:} $\epsilon^\mathrm{init}$, $\epsilon^{\mathrm{cool}}$, $\zeta^{\mathrm{tol}}$, $k^{\mathrm{max}}$
\State \textbf{set:} Calculate Sobol grid of starting grains $(R_n, x_n)_{n=1}^{N^{\mathrm{s}}}$
\State \textbf{set:} $L_{\cdot 0} \leftarrow L^{\rm{out}}$
\State \textbf{set:} $L_{0 \cdot} \leftarrow = \sigma^{-2} - \log(4 \pi \sigma^2 \sinh (\sigma^{-2}))$ 

\For{$0<n<N^{\mathrm{s}}$}
  
  \vspace{0.5em}
  \LineComment{\emph{Use a sample with a single grain}}

  \State $\epsilon \leftarrow \epsilon^\mathrm{init}$

  \vspace{0.5em}
  \For{ $ 0 < k < k^{\mathrm{max}}$ }

    \vspace{0.5em}
    \State Calculate model rays, $\forall \ j \in J$:
    \State $r_j \leftarrow (\mathbb{I} - 2 \Gamma_{t_j} R^{k}_{n_j} w_{m_j} w_{m_j}\T R^{k\top}_{n_j} \Gamma_{t_j}\T )e$
    \vspace{0.5em}
    \State Calculate detected ray estimates, $\forall \ i \in I, \ j \in J$:
    \State $\hat r_{ij} \leftarrow \frac{s_i - \Gamma_{t_j} x^{k}_{n_j}}{\| s_i - \Gamma_{t_j} x^{k}_{n_j} \|}$
    
    \vspace{0.5em}
    \State Log likelihood matrix, $\forall \ i \in I, \ j \in J$ :
    \State $L_{ij} \leftarrow \frac{1}{\sigma^2} \hat r_{ij}\T r_j$

      \vspace{0.5em}
      \LineComment{\emph{E-step}}
      \vspace{0.5em}
      \State Update assignments: 
      \State $K \leftarrow \exp(L/\epsilon)$
      \State $Q_n^{k} \leftarrow K \cdot \mathbf{diag}(a \oslash K \mathbbm{1})$

      \vspace{0.5em}
      {\LineComment{\emph{M-step}}}
      \vspace{0.5em}

      \State Update positions and orientations, $\ \forall \ n \in N:$
      \State $x_n^{k+1} \leftarrow \mathbf{LauePosNewton}[x^{k}, Q_n^{k}, ... ]$
      \State $R_n^{k+1} \leftarrow \mathbf{LaueRotMM}[R^k_n, Q_n^{k}, ...]$

      \vspace{0.5em}
      \State Decrease the temperature: $\epsilon \leftarrow \epsilon\cdot\epsilon^{\mathrm{cool}}$
      \State $k \leftarrow k+1$
      \vspace{0.5em}

  \EndFor

\vspace{0.5em}
\EndFor
\State \textbf{output:} $(R_n, x_n)_{n=1}^{N^{\mathrm{s}}}$
\end{algorithmic}
\end{algorithm}

\begin{algorithm}
\caption{\textbf{LaueOT}: Laue indexing with optimal transport, multi-grain fitting}
\label{alg:multigrain}
\begin{algorithmic}[1]
\footnotesize
\State \textbf{input data:} $(s_i)_{i\in I}$, $\sigma$  
\State \textbf{input prototype grain parameters:} $(R^{0}_n,x^{0}_n)_{n \in N}$ 
\State \textbf{input OT parameters:} $L^{\rm{out}}$, $T_a$
\State \textbf{input control parameters:} $\epsilon^\mathrm{init}$, $\epsilon^{\mathrm{cool}}$, $\zeta^{\mathrm{tol}}$

\State \textbf{set:} $L_{0 \cdot} \leftarrow = \sigma^{-2} - \log(4 \pi \sigma^2 \sinh (\sigma^{-2}))$, 
\State \textbf{set:} $L_{\cdot0} \leftarrow L^{\rm{out}}$
\State \textbf{set:} $\epsilon \leftarrow \epsilon^\mathrm{init}$, $\zeta \leftarrow \inf$
\While{ $\epsilon{\geq}1$ or $\zeta > \zeta^{\rm{tol}}$  }

    \vspace{0.5em}
    \State Calculate model rays, $\forall \ j \in J$:
    \State $r_j \leftarrow (\mathbb{I} - 2 \Gamma_{t_j} R^{k}_{n_j} w_{m_j} w_{m_j}\T R^{k \top}_{n_j} \Gamma_{t_j}\T )e$
    \vspace{0.5em}
    \State Calculate detected ray estimates, $\forall \ i \in I, \ j \in J$:
    \State $\hat r_{ij} \leftarrow \frac{s_i - \Gamma_{t_j} x^{k}_{n_j}}{\| s_i - \Gamma_{t_j} x^{k}_{n_j} \|}$
    
    \vspace{0.5em}
    \State Calculate log-likelihood matrix, $\forall \ i \in I, \ j \in J$ :
    \State $L_{ij} \leftarrow \frac{1}{\sigma^2} \hat r_{ij}\T r_j$

    \vspace{1em}
    \LineComment{\emph{E-step}}
    \State Update assignments by solving OT:
    \State $Q^{k+1} \leftarrow  \mathbf{BalancedOT}[L, \epsilon, ...]$

    \vspace{0.5em}
    \LineComment{\emph{M-step}} 

    \State Update orientations and positions, $\ \forall \ n \in N:$
    \State $x_n^{k+1} \leftarrow \mathbf{LauePosNewton}[x^{k}_n, Q^{k+1}, ... ]$
    \State $R_n^{k+1} \leftarrow \mathbf{LaueRotMM}[R^k_n, Q^{k+1}, ...]$
        
    \vspace{0.5em}
    \LineComment{\emph{Deterministic annealing temperature control}}
    \If{$\epsilon > 1$}
      \State Decrease temperature: $\epsilon \leftarrow \epsilon\cdot\epsilon^{\mathrm{cool}}$
    \EndIf

    \vspace{0.5em}
    \State $\zeta \leftarrow \|Q^{k+1}-Q^{k}\|_F $
    \State $k \leftarrow k+1$
    \vspace{0.5em}

\EndWhile
\State \textbf{output:} $Q^{k}$, $(R^{k}_n, x^{k}_n)_{n \in N}$
\end{algorithmic}
\end{algorithm}

\newpage
\ifCLASSOPTIONcompsoc
  \section*{Acknowledgments}
\else
  \section*{Acknowledgment}
\fi

The authors thank Marc Caubet Serrabou and administrators of the Merlin cluster at PSI for excellent technical support.
We thank Andreas Adelmann for the access to the Gwendolen node.
TK thanks Luis Barba for helpful discussions and Ilnura Usmanova for valuable comments on the manuscript.
The authors thank Benjamin Bejar for his initial work conceptual works and contributions to proposals that led to this project.
The project was supported by grant C20-14 from Swiss Data Science Center. 

\ifCLASSOPTIONcaptionsoff
  \newpage
\fi

\bibliographystyle{IEEEtran}
\bibliography{IEEEabrv,references}

\end{document}